\documentclass[useAMS,usenatbib]{mn2e}
\usepackage{natbib}
\usepackage{epsfig}
\usepackage{longtable}
\usepackage{times}
\title[INTEGRAL results on SFXTs]{\emph{INTEGRAL} results on Supergiant Fast X$-$ray Transients and accretion mechanism 
interpretation: ionization effect and formation of transient accretion disks}
\author[L. Ducci, L. Sidoli and A. Paizis]{L. Ducci,$^{1,2}$\thanks{E-mail: lorenzo@iasf-milano.inaf.it} L. Sidoli,$^{2}$ A. Paizis$^{2}$\\
$^{1}$ Dipartimento di Fisica e Matematica, Universit\`a degli Studi dell'Insubria, Via Valleggio 11, I-22100 Como, Italy \\
$^{2}$ INAF, Istituto di Astrofisica Spaziale e Fisica Cosmica, Via E. Bassini 15, I-20133 Milano, Italy}
\begin{document}

\pagerange{\pageref{firstpage}--\pageref{lastpage}} \pubyear{2010}

\maketitle

\label{firstpage}

\begin{abstract}
We performed a systematic analysis of all \emph{INTEGRAL} observations
from 2003 to 2009 of 14 Supergiant Fast X$-$ray Transients (SFXTs),
implying a net exposure time of about $30$~Ms.
For each source we obtained lightcurves and spectra (3$-$100~keV),
discovering several new outbursts.
We discuss the X$-$ray behaviour of SFXTs emerging from our analysis
in the framework of the clumpy wind accretion mechanism 
we proposed \citep{Ducci2009}.
We discuss the effect of X$-$ray photoionization on accretion in close binary systems
like IGR~J16479$-$4514 and IGR~J17544$-$2619.
We show that, because of X$-$ray photoionization,
there is a high probability of formation of an accretion disk
from capture of angular momentum in IGR~J16479$-$4514, 
and we suggest that the formation of transient accretion disks could be
responsible of part of the flaring activity in SFXTs with narrow orbits.
We also propose an alternative way to explain 
the origin of flares with peculiar shapes observed in our analysis
applying the model of \citet{Lamb1977}, which is
based on the accretion via Rayleigh$-$Taylor instability,
and was originally proposed to explain type II bursts.
\end{abstract}

\begin{keywords}
X--rays: binaries -- X-rays: individuals: IGR J16479--4514, XTE J1739--302/IGR J17391--3021, AX J1841.0--0536/IGR J18410--0535,
IGR J18483--0311, SAX J1818.6--1703, IGR J16418--4532, AX J1820.5--1434,   
AX J1845.0--0433, IGR J16195--4945, IGR J16207--5129, IGR J16465--4507, IGR J17407--2808, XTE J1743--363.     
\end{keywords}

\section{Introduction}
\label{Introduction}

Supergiant Fast X$-$ray Transients (SFXTs)
are a sub-class of High Mass X$-$ray Binaries (HMXBs)
discovered by the \emph{INTEGRAL} satellite
in the last seven years, during the Galactic plane monitoring \citep{Sguera2005}.
SFXTs host an OB supergiant and an accreting compact object,
and show a sporadically X$-$ray transient emission composed by many flares
reaching a luminosity of $10^{36}-10^{37}$~erg~s$^{-1}$,
with flare durations of $\sim 10^3-10^4$~s.
For most of their lifetime, SFXTs accrete at an intermediate level, 
showing an X$-$ray luminosity of $10^{33}-10^{34}$~erg~s$^{-1}$, as discovered by
the \emph{Swift}/XRT monitoring (see e.g. \citealt{Sidoli2008}; \citealt{Romano2009}).
The quiescent level (about $10^{32}$~erg~s$^{-1}$)
has been observed only in a few sources 
(see \citealt{intZand2005}; \citealt{Leyder2007}; \citealt{Bozzo2010}).
Thus SFXTs show a large dynamic range of about $10^3-10^5$.
Many accretion mechanisms have been proposed to explain the transient
behaviour. \citet{intZand2005} suggested that the large dynamic range
could be produced by the accretion of dense clumps from the donor wind.
\citet{Sidoli2007} invoked the presence of an equatorial wind component,
denser than the polar wind component and 
inclined with respect to the orbital plane of the compact object
to explain the transient periodic emission of IGR~J11215-5952.
\citet*{Bozzo2008a} proposed that SFXTs host a magnetar with large spin period ($\sim 10^3$~s):
the changes in X$-$ray luminosity are ascribed to 
the gated accretion mechanisms, where changes in the reciprocal positions of 
accretion, magnetospheric and corotation radii lead to different accretion regimes.

In this paper we report the results from 
the analysis of \emph{INTEGRAL} data
of 14 SFXTs, for a total exposure time of $\sim 30$~Ms.
The results obtained here are discussed considering
the structure of the clumpy supergiant winds \citep{Ducci2009},
the effect of X$-$ray photoionization of the outflowing wind
by the compact object (in the framework of the \citealt{Ho1987} accretion model), 
and the possible formation of transient accretion disks,
as those proposed by \citet*{Taam1988} to reproduce the flares from EXO~2030+375.

\section{Observations and Data analysis}
\label{Data analysis}

\begin{figure*}
\begin{center}
\includegraphics[width=8cm,bb = 100 360 558 720,clip]{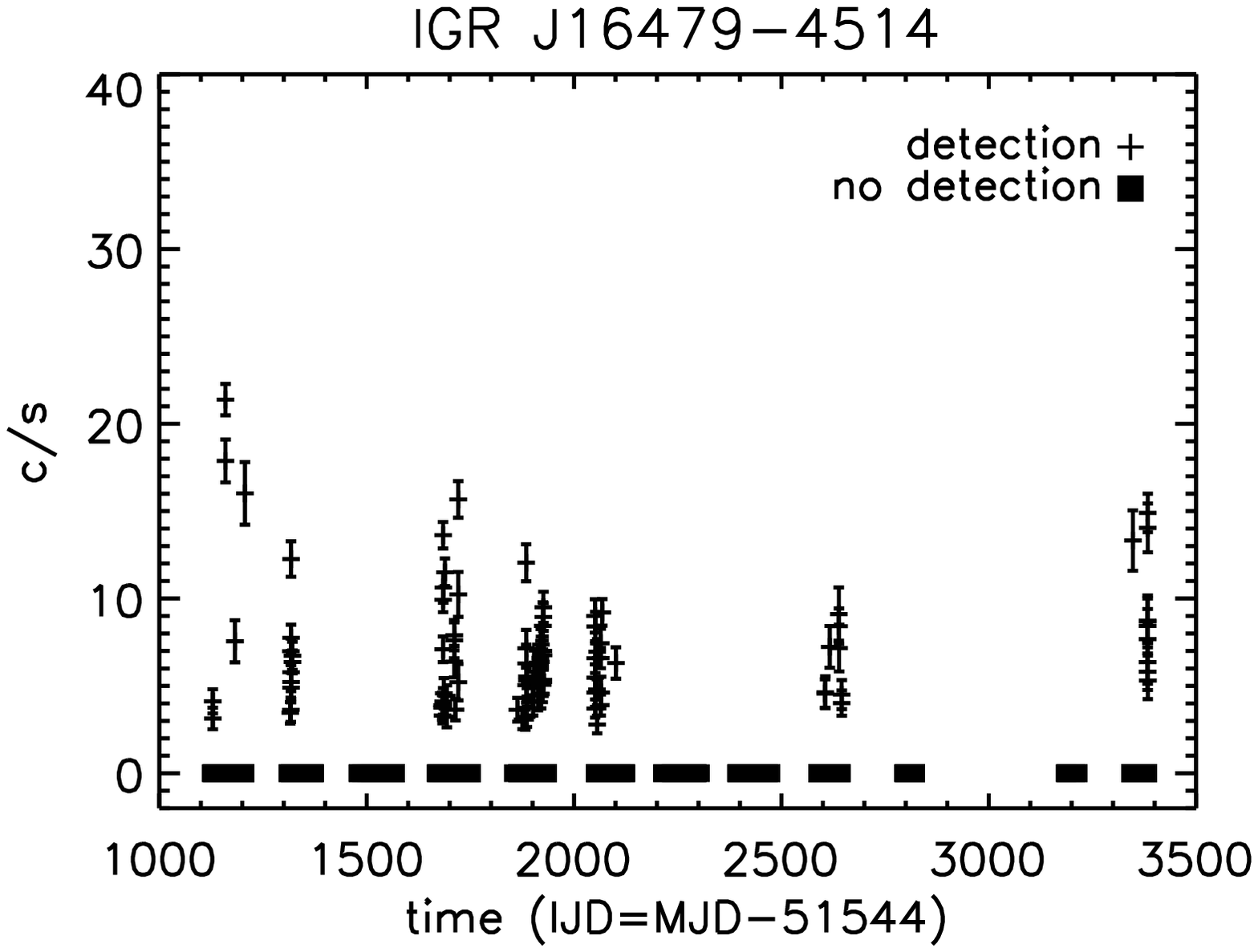}
\includegraphics[width=8cm,bb = 100 360 558 720,clip]{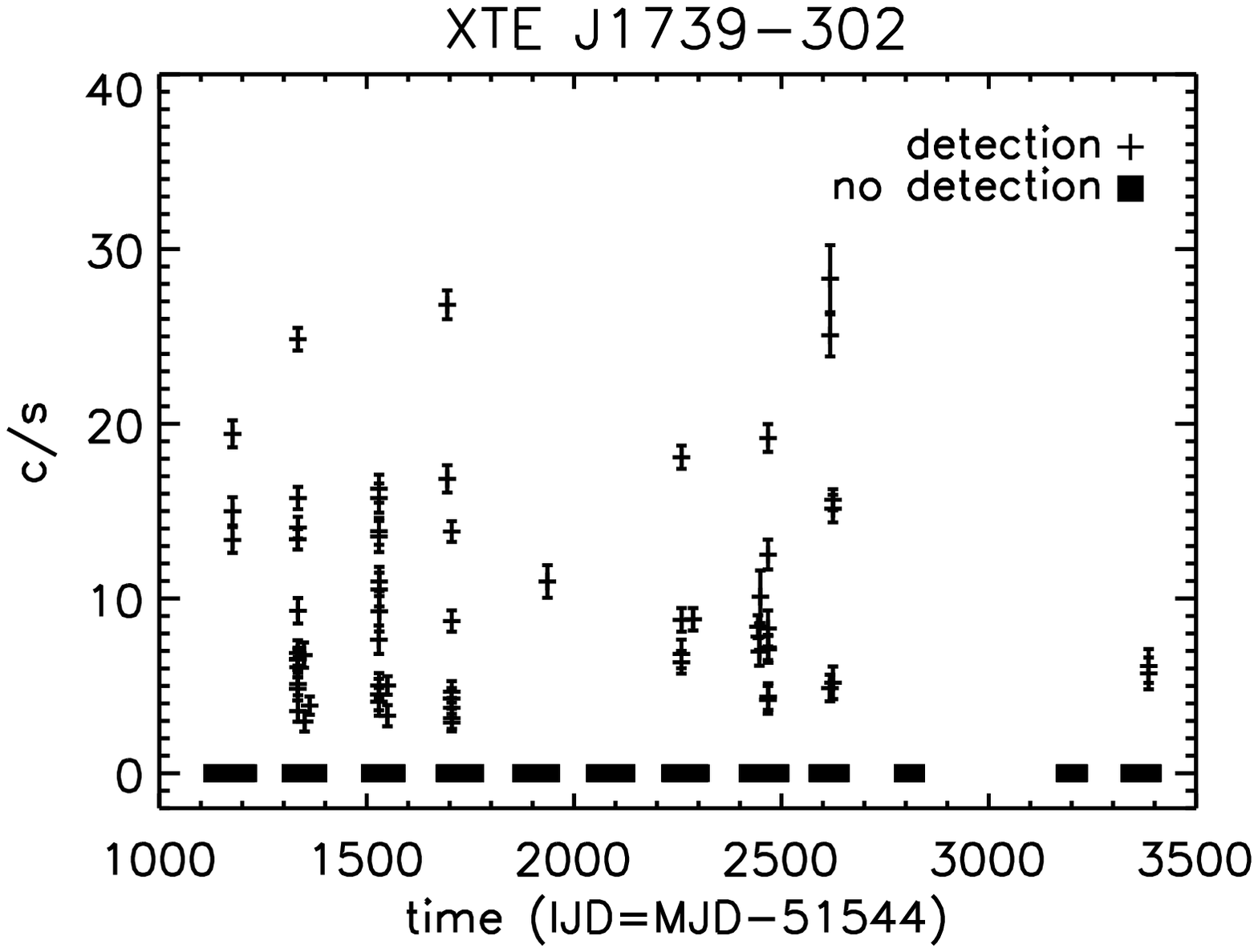}\\
\includegraphics[width=8cm,bb = 100 360 558 720,clip]{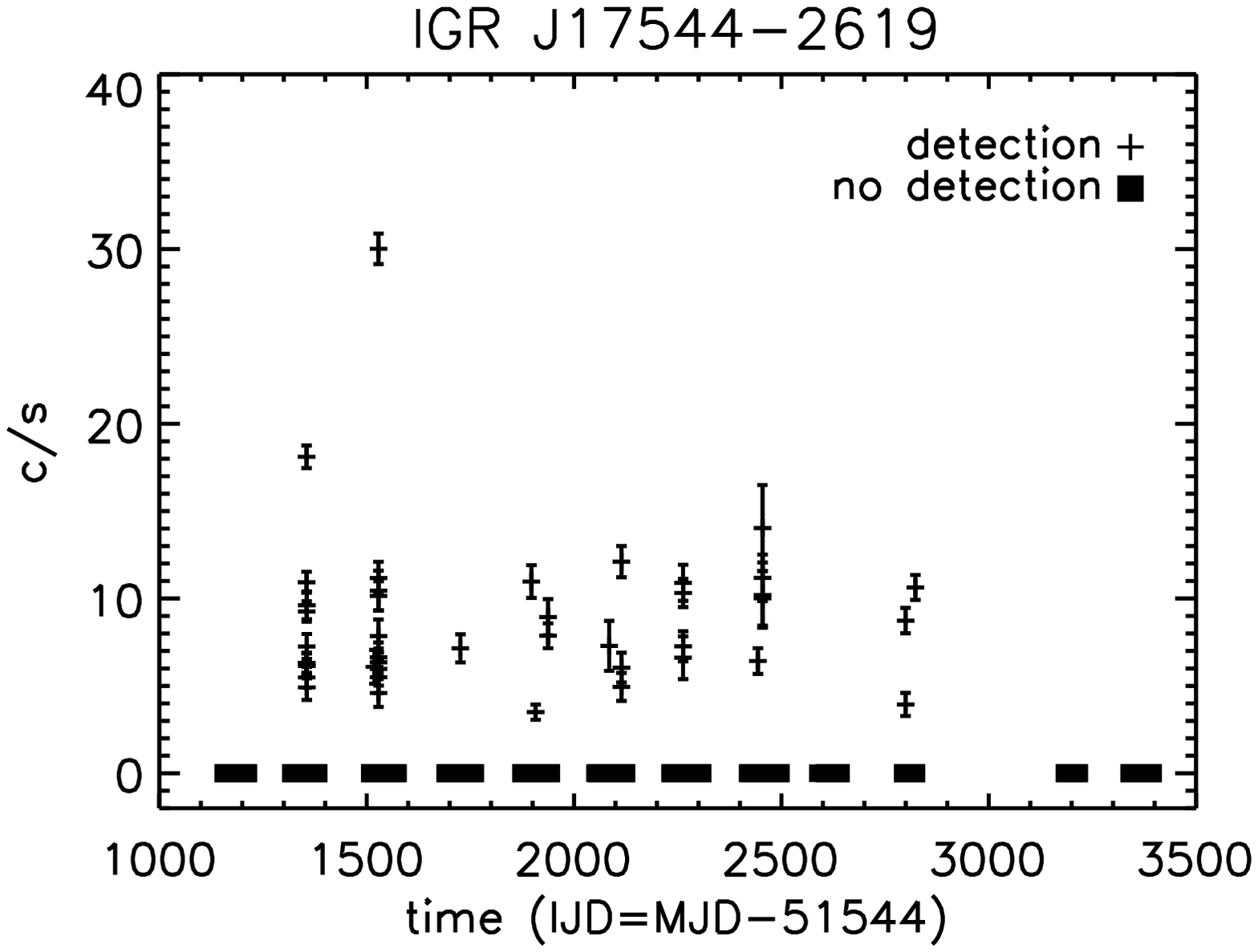}
\includegraphics[width=8cm,bb = 100 360 558 720,clip]{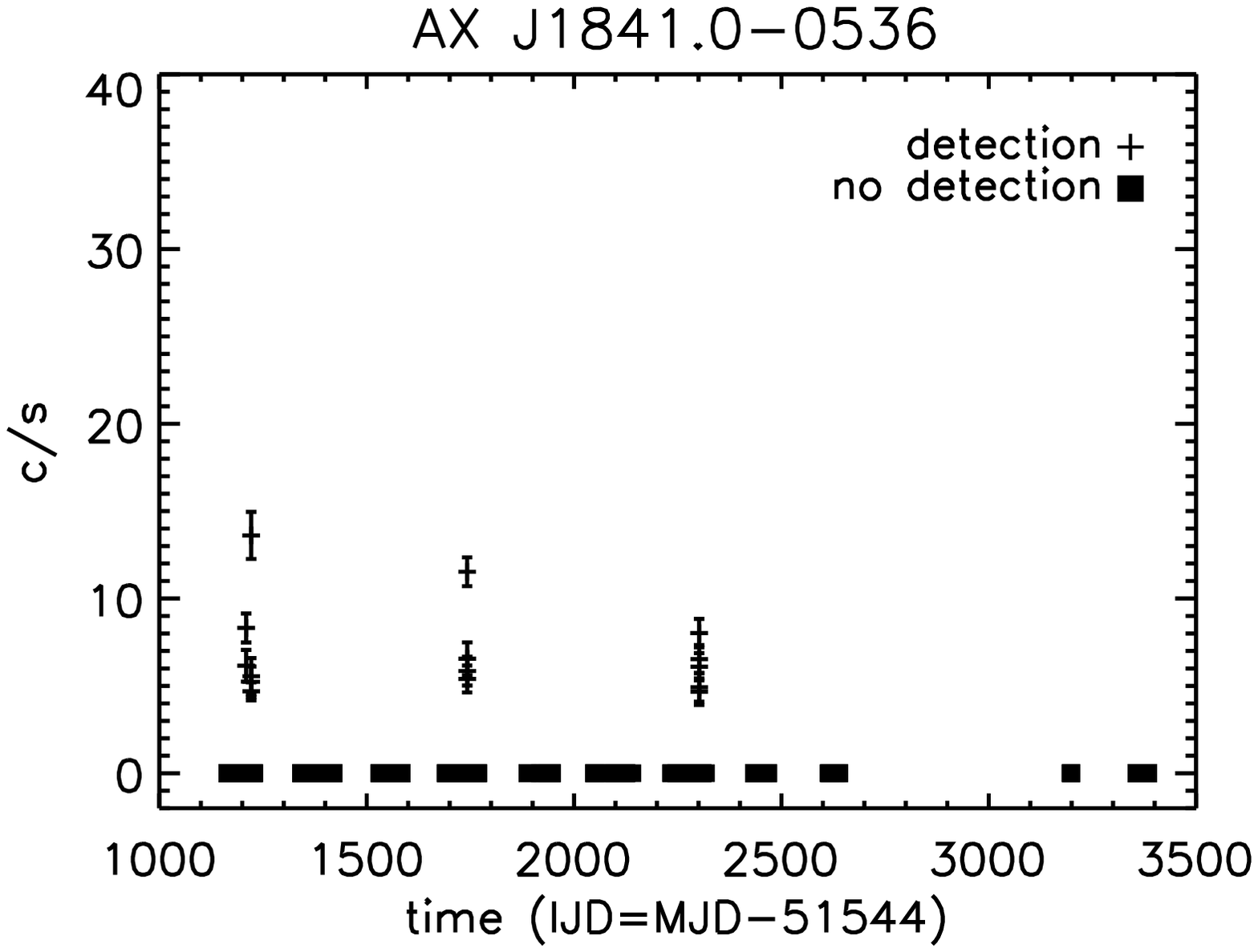}
\end{center}
\caption{IBIS/ISGRI lightcurves of IGR~J16479$-$4514, XTE~J1739$-$302, IGR~J17544$-$2619, AX~J1841.0$-$0536 ($20-40$~keV). 
A binned time corresponding to the ScW duration ($\sim2000$~s) has been used.} 
\label{lcr_tot_igrj16479.ps}
\end{figure*}

\begin{table}
\begin{center}
\caption{Total exposure time $T_{\rm exp}$, duration and number of ScWs 
where the source is detected with a significance greater than 5.}
\label{table Texp}
\begin{tabular}{lcccc}
\hline
Source            & $T_{\rm exp}$ (d) &  Duration (d)& \# ScWs &  \# Outbursts  \\
                  &                  &  $S/N>5$     & $S/N>5$ &               \\
\hline
IGR J16479-4514   &     115.74      &     3.19      &  107    &       38      \\
XTE J1739-302     &     260.42      &     2.16      &   65    &       18      \\
IGR J17544-2619   &     259.92      &     1.33      &   43    &       14      \\
AX J1841.0-0536   &      83.23      &     0.41      &   16    &        4      \\
IGR J18483-0311   &      84.61      &     2.68      &   90    &       13      \\
SAX J1818.6-1703  &     179.45      &     1.08      &   34    &       11      \\
IGR J16418-4532   &     112.87      &     1.45      &   40    &       23      \\
AX J1820.5-1434   &     120.95      &     0.09      &    4    &        2     \\
AX J1845.0-0433   &      77.85      &     0.21      &    7    &        7     \\
IGR J16195-4945   &     105.22      &     0.14      &    4    &        3     \\
IGR J16207-5129   &     101.67      &     0.48      &   15    &        9     \\
IGR J16465-4507   &     110.25      &     0.27      &   10    &        2     \\
IGR J17407-2808   &     234.75      &     0.12      &    4    &        3     \\
XTE J1743-363     &     232.04      &     0.48      &   20    &        7     \\
\hline
\end{tabular}
\end{center}
\end{table}

The \emph{INTEGRAL} observatory, launched in October 2002,
carries 3 co-aligned coded mask telescopes:
the imager IBIS (Imager on Board the \emph{INTEGRAL} satellite;
\citealt{Ubertini2003}), sensitive from 15~keV to 10~MeV,
the spectrometer SPI (SPectrometer on \emph{INTEGRAL}, 
20~keV$-$8~MeV; \citealt{Vedrenne2003}),
and the two X$-$ray monitors JEM-X1 and JEM-X2 (Joint European X$-$ray Monitor; 
\citealt{Lund2003}), sensitive in the energy range 3$-$35~keV.
IBIS is composed of the low-energy detector ISGRI 
(\emph{INTEGRAL} Soft Gamma Ray Instrument; 15$-$600~keV; \citealt{Lebrun2003})
and the CsI layer PICsIT 
(Pixellated Imaging Caesium Iodide Telescope; 175~keV$-$10~MeV; \citealt{Labanti2003}).
\emph{INTEGRAL} observations are divided in pointings
called Science Windows (ScWs), which have a typical exposure of 2~ks.

We analysed all the public and our private data,
between 2003 and 2009, where the SFXTs IGR~J16479$-$4514,
XTE~J1739$-$302, AX~J1841.0$-$0536 and IGR~J17544$-$2619 
were within 15$^\circ$ from the center of the field of view.
This resulted in 14426 Science Windows,
corresponding to a total exposure time of $\sim 30$~Ms.
We considered in our analysis also the other SFXTs
(and candidate SFXTs) observed in the ScWs selected (see Table \ref{table Texp}). 
We analysed IBIS/ISGRI and JEM-X data using the Off-line Scientific Analysis 
package OSA 8.0 \citep{Goldwurm2003}.
For the spectral analysis, which was performed with {\scshape xspec} (ver. 11.3),
we added a 2\% systematic error to both IBIS/ISGRI and JEM-X data sets.

\begin{figure*}
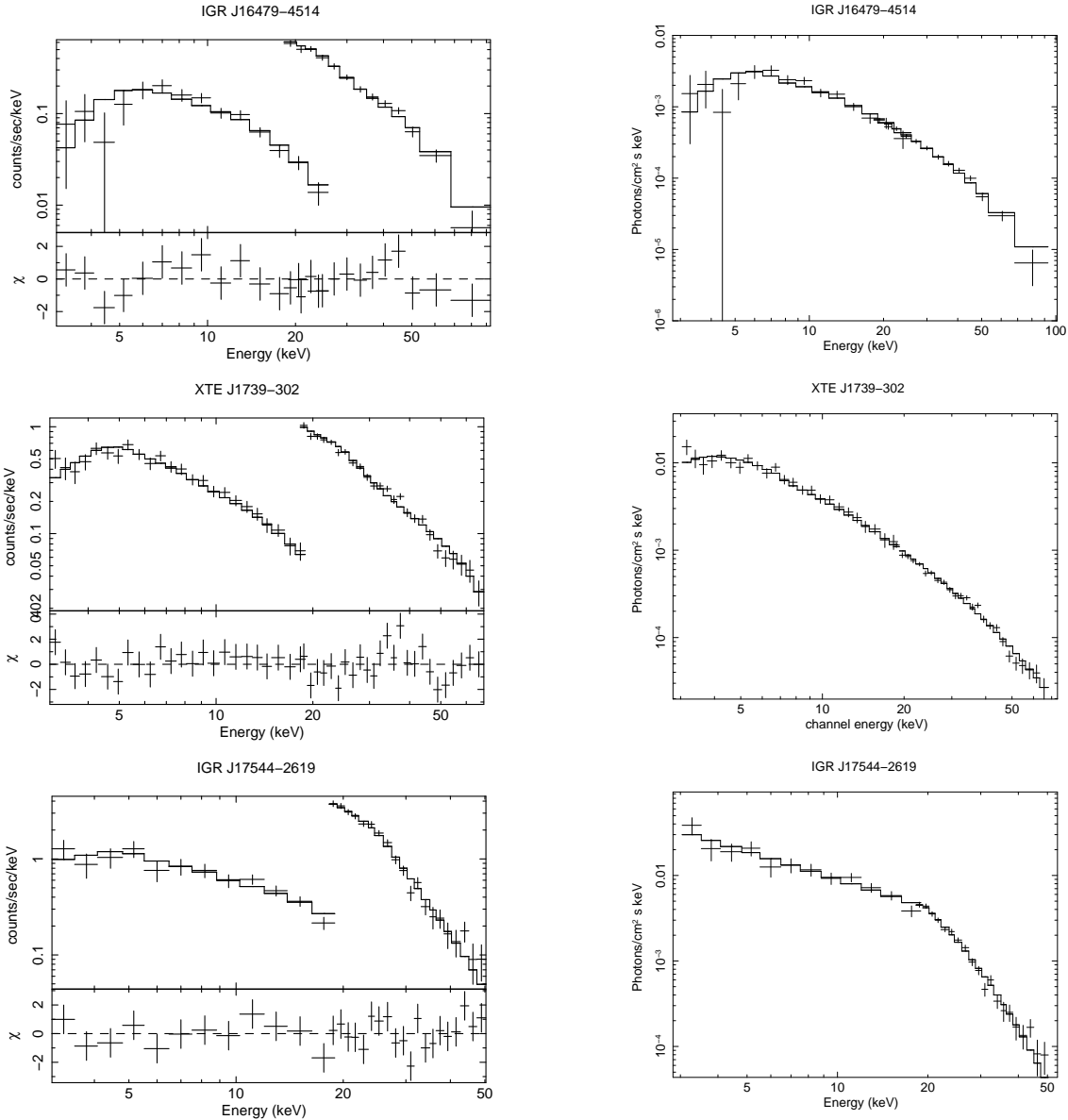

\begin{center}
\includegraphics[width=5.3cm, angle=-90]{igrj16479_mosaico_wabs_cutoffpl.ps}
\includegraphics[width=5cm, angle=-90]{ufs_igrj16479_mosaico_wabs_cutoffpl.ps}
\includegraphics[width=5.3cm, angle=-90]{xtej1739_jemx_isgri_po_cutoffpl.ps}
\includegraphics[width=5cm, angle=-90]{ufs_xtej1739_jemx_isgri_po_cutoffpl.ps}
\includegraphics[width=5.3cm, angle=-90]{igrj17544_jemx_isgri_phabs_po_highecut.ps}
\includegraphics[width=5cm, angle=-90]{ufs_igrj17544_jemx_isgri_phabs_po_highecut.ps}
\end{center}
\caption{Joint JEM-X plus IBIS/ISGRI counts spectra, together with residuals in units of standard deviations 
(left panels) and unfolded spectra (right panels)
for IGR~J16479$-$4514, XTE~J1739$-$302, and IGR~J17544$-$2619 (see Table \ref{table par. spettri jemx}).}
\label{figura spettri jemx}
\end{figure*}

For each source reported in Table \ref{table Texp}, 
we extracted the IBIS/ISGRI lightcurve at a ScW time 
resolution ($\sim 2000$~s), in the energy range 20$-$40~keV,
and we considered only the pointings where the sources 
are detected with at least a $5\sigma$ significance.
For each source, Table \ref{table Texp} reports  
the number of outbursts\footnote{We defined \emph{outburst} as an X$-$ray emission
detected by IBIS/ISGRI with significance $>5\sigma$, and separated by
adjacent outbursts by at least $\sim$1~day of inactivity 
(source below the IBIS/ISGRI threshold of detectability).
An outburst can be composed by one or a series of flares
with a typical duration of $\sim 10^2-10^4$~s each.} 
we identified.
For each outburst observed, 
we also extracted IBIS/ISGRI  
lightcurves (18$-$60~keV) with a bin time of 50~s, and an IBIS/ISGRI spectrum
in the energy range 18$-$100~keV.
We found that the outbursts are characterized by a flaring activity,
typical of SFXTs.
Due to its smaller field of view,
we were able to extract the JEM-X spectra 
in only a few pointings.

\section{Results}
\label{Results}

With IBIS/ISGRI we discovered several previously unnoticed outbursts 
from XTE~J1739$-$302, IGR~J16479$-$4514, IGR~J17544$-$2619,
IGR~J18483$-$0311, AX~J1845.0$-$0433, IGR~J16195$-$4945,
IGR~J16465$-$4507, IGR~J17407$-$2808, IGR~J16207$-$5129, 
and IGR~J16418$-$4532. We reported these outbursts 
in Table \ref{table nuovi outbursts} 
and \ref{table nuovi outbursts candidate}, together with the 
mean flux and best fit parameters of the IBIS/ISGRI spectra.

Figure \ref{lcr_tot_igrj16479.ps} displays
the 20$-$40~keV IBIS/ISGRI lightcurves of 
IGR~J16479$-$4514, XTE~J1739$-$302,
IGR~J17544$-$2619 and AX~J1841.0$-$0536
collected from 2003 to 2009,
where each ``detection'' refers to the 
average flux observed during each ScW,
when the sources are detected at a significance $>5\sigma$.
Solid boxes represent the
observations where the sources are not detected ($<5\sigma$).
For most of the time each source is not significantly detected.

\begin{table*}
\begin{center}
\caption{Summary of the new flares discovered of 7 confirmed SFXTs (IBIS/ISGRI data).
We fitted the spectra with a powerlaw or a bremsstrahlung model.}
\label{table nuovi outbursts}
\begin{tabular}{llccc}
\hline
Source           &                   Date             &       Mean Flux           &     $kT$ (keV)/$\Gamma$     &$\chi^2_{\nu}$ (d.o.f.)\\
                 &                  (UTC)             &   (erg~cm$^{-2}$~s$^{-1}$) &                             &                      \\
\hline
XTE~J1739$-$302    &    2009 Apr. 8, 16:49$-$18:46      &    $4 \times 10^{-10}$     &$\Gamma=2.8{+0.5 \atop -0.5}$&  $0.99$ ($10$)  \\
\hline
IGR J16479-4514  &    2003 Feb.  2  19:07$-$22:30     &    $2.5 \times 10^{-10}$   &$\Gamma=2.6{+0.5 \atop -0.5}$&  $0.76$ ($9$)   \\
                 &    2004 Aug.  9, 02:26$-$04:24     &    $3.0 \times 10^{-10}$   &    $kT=41{+45 \atop -17}$   &  $0.52$ ($13$)  \\
                 &    2004 Aug. 20, 07:27$-$12:58     &    $2.6 \times 10^{-10}$   &$\Gamma=2.1{+0.5 \atop -0.5}$&  $0.73$ ($10$)  \\
                 &    2004 Sep. 10, 01:14$-$02:12     &    $2.5 \times 10^{-10}$   &    $kT=19{+19 \atop -8}$    &  $0.4$ ($12$)   \\ 
                 &    2009 Mar.  1, 16:56$-$17:54     &         $10^{-9}$          &$\Gamma=2.7{+0.7 \atop -0.6}$&  $1.2$ ($9$)    \\
                 &    2009 Apr.  6,  6:16$-$13:09     &    $7.2 \times 10^{-10}$   &      $kT=34{+7 \atop -5}$   &  $1.45$ ($12$)  \\
\hline
IGR J17544-2619  &    2004 Feb. 27, 14:18$-$14:45     &    $4.2 \times 10^{-10}$   &$\Gamma=3.1{+0.9 \atop -0.8}$&  $1.27$ ($8$)   \\
                 &    2006 Sep. 20, 09:58$-$13:47     &    $6.9 \times 10^{-10}$   &    $kT=18{+29 \atop -8}$    &  $1.38$ ($7$)   \\ 
\hline
IGR J18483-0311  &    2004 Nov.  3, 00:05$-$00:31     &    $3.3 \times 10^{-10}$   &    $kT=22{+48 \atop -11}$   &  $1.01$ ($11$)  \\ 
                 &    2005 Oct. 16, 07:34$-$08:30     &    $1.7 \times 10^{-10}$   &    $kT=21{+39 \atop -10}$   &  $1.19$ ($11$)  \\ 
                 &    2005 Oct. 19, 20:48$-$20, 18:40 &    $3.1 \times 10^{-10}$   &$\Gamma=2.8{+0.3 \atop -0.3}$&  $1.42$ ($10$)  \\
                 &    2006 Apr. 25, 14:04$-$14:33     &    $5.0 \times 10^{-10}$   &    $kT=40{+21 \atop -21}$   &  $1.54$ ($11$)  \\ 
\hline
AX J1845.0-0433  &    2006 Sep.  3, 17:52$-$18:28     &    $4.6 \times 10^{-10}$   &$\Gamma=2.4{+0.5 \atop -0.5}$&  $0.83$ ($11$)  \\
\hline
IGR J16195-4945  &    2004 Aug. 20, 04:40$-$06:27     &    $2.6 \times 10^{-10}$   &$\Gamma=2.0{+0.6 \atop -0.6}$&  $0.73$ ($10$)  \\
                 &    2005 Feb. 18, 13:25$-$14:22     &    $2.6 \times 10^{-10}$   &$\Gamma=2.0{+1.1 \atop -0.9}$&  $0.61$ ($8$)   \\
\hline
IGR J16465-4507  &    2004 Aug.  9, 22:12$-$10, 02:53 &    $2.3 \times 10^{-10}$   &$\Gamma=2.5{+0.5 \atop -0.5}$&  $0.86$ ($10$)  \\
\hline
\end{tabular}
\end{center}
\end{table*}

\begin{table*}
\begin{center}
\caption{Summary of the new flares discovered of 3 candidate SFXTs (IBIS/ISGRI data). 
We fitted the spectra with a powerlaw or a bremsstrahlung model.}
\label{table nuovi outbursts candidate}
\begin{tabular}{llccc}
\hline
Source           &              Date                  &     Mean Flux           &       $kT$ (keV)/$\Gamma$     &$\chi^2_{\nu}$ (d.o.f.)\\
                 &              (UTC)                 & (erg~cm$^{-2}$~s$^{-1}$) &                               &                      \\
\hline
IGR J17407-2808  &    2003 Sep. 21, 04:27$-$04:29     &  $2.6\times 10^{-9}$     &     $kT=44{+35 \atop -16}$   &  $1.38$ ($10$) \\
                 &    2006 Sep. 20, 12:18$-$13:47     &  $4.9 \times 10^{-10}$   &$\Gamma=1.9{+0.3 \atop -0.3}$ &  $0.55$ ($14$) \\
\hline
IGR J16207-5129  &    2004 Jan. 21, 09:23$-$09:56     &  $5.8 \times 10^{-10}$   &    $kT=38.{+81 \atop -17}$   &  $1.37$ ($10$) \\ 
                 &    2005 Feb. 07, 06:48$-$07:45     &  $2.0 \times 10^{-10}$   &     $kT=30{+94 \atop -16}$   &  $0.69$ ($11$) \\ 
                 &    2005 Feb. 11, 03:04$-$04:01     &  $2.2 \times 10^{-10}$   &$\Gamma=2.0{+0.9 \atop -0.8}$ &  $0.78$ ($10$) \\
                 &    2005 Feb. 13, 10:59$-$11:56     &  $2.2 \times 10^{-10}$   &$\Gamma=1.8{+0.8 \atop -0.8}$ &  $1.06$ ($10$) \\
                 &    2005 Feb. 18, 22:41$-$19, 01:41 &  $3.2 \times 10^{-10}$   &$\Gamma=2.2{+0.3 \atop -0.3}$ &  $1.06$ ($11$) \\
\hline
IGR J16418-4532  &    2003 Feb.  3, 12:37$-$13:13     &  $4.6 \times 10^{-10}$   &$\Gamma=1.8{+1.2 \atop -1.1}$ &  $1.07$ ($5$)  \\
                 &    2003 Mar.  4, 21:39$-$22:07     &  $2.1\times 10^{-10}$    &     $kT=15{+16 \atop -6}$    &  $1.12$ ($14$) \\
                 &    2004 Feb. 18, 03:42$-$04:16     &  $3.0 \times 10^{-10}$   &$\Gamma=3.2{+1.2 \atop -1.0}$ &  $1.75$ ($14$) \\
                 &    2004 Mar. 20, 21:27$-$21, 11:24 &  $2.5 \times 10^{-10}$   &$\Gamma=3.0{+0.7 \atop -0.6}$ &  $0.96$ ($14$) \\
                 &    2004 Aug.  9, 13:57$-$14:56     &  $1.9 \times 10^{-10}$   &$\Gamma=2.7{+0.9 \atop -0.7}$ &  $0.70$ ($14$) \\
                 &    2004 Aug. 24, 13:32$-$14:56     &  $2.0\times 10^{-10}$    &     $kT=20{+12 \atop -7}$    &  $1.00$ ($14$) \\
                 &    2005 Feb. 13, 02:17$-$15:02     &  $1.6\times 10^{-10}$    &     $kT=15{+6 \atop -4}$     &  $1.34$ ($14$) \\
                 &    2005 Feb. 22, 04:54$-$06:14     &  $1.5\times 10^{-10}$    &     $kT=22{+23 \atop -10}$   &  $1.13$ ($14$) \\  
                 &    2005 Feb. 24, 05:20$-$07:10     &  $1.3 \times 10^{-10}$   &$\Gamma=2.8{+1.1 \atop -0.9}$ &  $0.87$ ($14$) \\
                 & 2005 Feb. 27, 05:41$-$Mar. 1, 12:09&  $1.9\times 10^{-10}$    &     $kT=21{+7 \atop -5}$     &  $1.02$ ($14$) \\  
                 &    2005 Mar.  8, 04:35$-$05:35     &  $3.2\times 10^{-10}$    &     $kT=14{+8 \atop -3}$     &  $1.16$ ($14$) \\  
                 &    2005 Mar. 18, 14:09$-$14:42     &  $1.3 \times 10^{-10}$   &$\Gamma=4.2{+5.8 \atop -1.4}$ &  $0.62$ ($7$)  \\
                 &    2005 Aug. 26, 18:08$-$27, 08:56 &  $2.5\times 10^{-10}$    &     $kT=28{+15 \atop -9}$    &  $1.16$ ($14$) \\  
                 &    2005 Oct.  2, 00:32$-$01:15     &  $3.6 \times 10^{-10}$   &$\Gamma=2.6{+1.6 \atop -1.4}$ &  $0.76$ ($8$)  \\
                 &    2006 Aug. 13, 23:48$-$14, 21:01 &  $2.9\times 10^{-10}$    &     $kT=20{+6 \atop -4}$     &  $1.65$ ($14$) \\ 
                 &    2007 Feb. 19, 03:58$-$04:39     &  $3.9 \times 10^{-10}$   &$\Gamma=2.3{+0.9 \atop -0.8}$ &  $0.74$ ($8$)  \\
                 &    2007 Mar. 15, 04:43$-$05:27     &  $5.4\times 10^{-10}$    &     $kT=15{+12 \atop -5}$    &  $1.48$ ($8$)  \\ 
                 &    2007 Mar. 30, 02:19$-$03:01     &  $3.0 \times 10^{-10}$   &$\Gamma=2.0{+1.9 \atop -1.6}$ &  $0.87$ ($5$)  \\
                 &    2009 Apr.  7, 10:15$-$10:43     &  $4.8 \times 10^{-10}$   & $\Gamma=2.8{+0.9 \atop -0.8}$&  $0.92$ ($10$) \\
                 &    2009 Apr.  7, 21:13$-$21:41     &  $4.1 \times 10^{-10}$   &$\Gamma=2.28{+0.99 \atop -0.91}$&$1.22$ ($9$)  \\
\hline
\end{tabular}
\end{center}
\end{table*}

\begin{figure}
\begin{center}
\includegraphics[height=7.5cm, angle=-90]{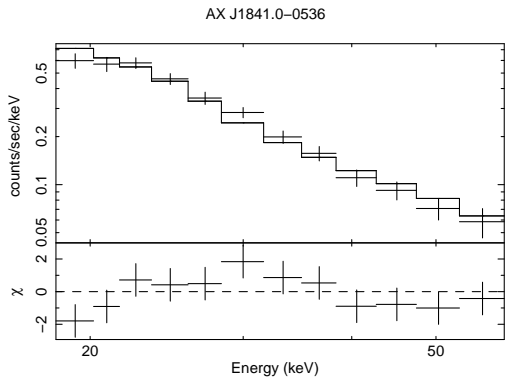}
\end{center}
\caption{Average IBIS/ISGRI counts spectra of AX~J1841.0$-$0536 
(see Table \ref{table par. spettri jemx}), and residuals (lowest panels)
in units of standard deviations.}
\label{figura spettri isgri}
\end{figure}

No evidence of spectral variability has been found
between different outbursts of these four SFXTs.
To achieve the best statistics, we extracted 
an average JEM-X$+$IBIS/ISGRI outburst spectrum 
for each source (Figures \ref{figura spettri jemx} and \ref{figura spettri isgri}).
We did not extract an average JEM-X spectrum for AX~J1841.0$-$0536 
because we detected the source only in one pointing, with a low flux.
The models which best fit the average spectra
are reported in Table \ref{table par. spettri jemx}.

\begin{table*}
\begin{center}
\caption{Best fit parameters of the average spectra of 
IGR~J16479$-$4514, XTE~J1739$-$302, and IGR~J17544$-$2619 
observed with JEM-X and IBIS/ISGRI (see Figure \ref{figura spettri jemx}),
and best fit parameters of the average spectrum of AX~J1841.0$-$0536
observed with IBIS/ISGRI (see Figure \ref{figura spettri isgri}).
$\Gamma$ is the powerlaw photon index, $E_{\rm c}$ is the cutoff energy, $E_{\rm F}$ is the e-folding energy.}
\label{table par. spettri jemx}
\begin{tabular}{lccccc}
\hline
Source          & Fit model &$N_{\rm H}$ ($10^{22}$~cm$^{-2}$)&           $\Gamma$          &$E_{\rm c}$, $E_{\rm F}$ (keV)&  $\chi^2_{\nu}$ (d.o.f.) \\
\hline
\multicolumn{6}{c}{JEM-X \& IBIS/ISGRI} \\
\hline
IGR J16479-4514 & cutoff-powerlaw & $23{+17 \atop -13}$   &    $1.2{+0.4 \atop -0.4}$   &$E_{\rm c}=25.1{+12 \atop -6}$&     $0.90$ (23)   \\
XTE J1739-302   & cutoff-powerlaw & $8.6{+4.1 \atop -3.6}$ &    $1.6{+0.2 \atop -0.2}$   &$E_{\rm c}=26.5{+6.5 \atop -4.7}$&     $1.12$ (45)   \\
IGR J17544-2619 & powerlaw with high-energy cutoff & $0.2{+8 \atop -0.15}$  &    $1.1{+0.3 \atop -0.1}$& $E_{\rm c}=19.7{+1.5 \atop -1.2}$; $E_{\rm F}=8.1{+1.0 \atop -0.7}$ &   $1.03$ (27) \\
\hline
\multicolumn{6}{c}{IBIS/ISGRI} \\
\hline
AX J1841.0-0536  &      powerlaw       & $-$ & $2.50{+0.16 \atop -0.15}$ &   $-$      &     $1.20$ (10)   \\
\hline
\end{tabular}
\end{center}
\end{table*}

\subsection{Clumpy wind in IGR~J16479$-$4514} \label{section clumpy wind}

The results of our analysis of all available \emph{INTEGRAL} observations of SFXTs
can be compared with quantitative expectations from our new clumpy
wind model \citep{Ducci2009}.
A meaningful comparison can be performed only in SFXTs where
a large number of flares has been observed.
For this reason we will concentrate only on IGR~J16479$-$4514.
In order to apply the model, we have to establish the stellar
parameters of the system and to confirm the eclipse duration.

\begin{figure}
\begin{center}
\includegraphics[height=6.2cm]{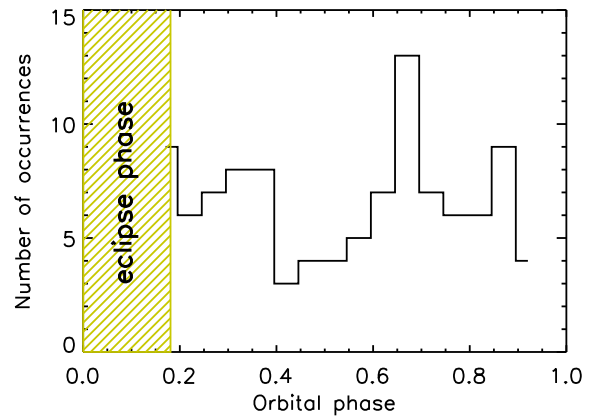}
\end{center}
\caption{Folding on a period of 3.3194~d of the ScWs where IGR~J16479$-$4514 is detected with 
significance $>5\sigma$.
The start time of the eclipse is $t_{\rm 0} = 54546.742$~MJD,
and the duration is 0.6~d.}
\label{istograma_scws_igrj16479}
\end{figure}
In Figure \ref{istograma_scws_igrj16479} we show the IBIS/ISGRI ScWs
where IGR~J16479$-$4514 has been detected in outburst,
folded on the orbital period of the system
assuming a zero time $t_{\rm 0} = 54546.742$~MJD \citep{Bozzo2008b}.
This histogram is consistent with the presence of an eclipse with duration 
$\Delta t \approx 0.6$~d  \citep*{Jain2009}.

\begin{figure}
\begin{center}
\includegraphics[bb= 100 372 535 700,clip, width=8cm]{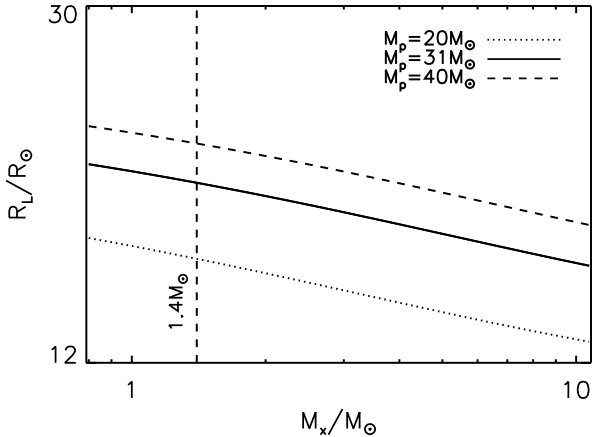}
\end{center}
\caption{$R_{\rm L}$ for different values of $q=M_{\rm p}/M_{\rm x}$
(see equation \ref{eggleton formula}).}
\label{eggleton figure}
\end{figure}

The Roche Lobe radius $R_{\rm L}$ of IGR~J16479$-$4514,
adopting the approximated formula obtained by \citet{Eggleton1983},
is the following:
\begin{equation} \label{eggleton formula}
 R_{\rm L} = a \frac{0.49q^{2/3}}{0.6q^{2/3} + \ln (1+q^{1/3})} \hspace{0.3cm} 0<q<\infty
\end{equation}
where $a$ is the orbital separation, $q=M_{\rm p}/M_{\rm x}$ 
is the mass ratio\footnote{$M_{\rm p}$ is the mass of the primary star (i.e. the supergiant star),
and $M_{\rm x}$ is the mass of the compact object.},
with a circular orbit assumed.
Figure \ref{eggleton figure}
reports the results obtained for $R_{\rm L}$.
Roche Lobe overflow (RLO) is not expected in this system because 
RLO would imply a much higher mass transfer to the compact object
and thus a much higher X$-$ray luminosity.
Assuming that the compact object is a neutron star with $M_{\rm x}=1.4$~M$_\odot$,
we make the hypothesis that the primary star has mass $M_{\rm p}=31$~M$_\odot$
and radius $R_{\rm p}=19$~R$_\odot$. 
This set of parameters is in agreement with the
expected values for a O8.5I star (e.g. \citealt*{Martins2005}; \citealt*{Vacca1996}),
and excludes the Roche Lobe overflow.
However, we point out that other parameters for $M_{\rm x}$, $M_{\rm p}$, and $R_{\rm p}$
are allowed.  

We applied our clumpy wind model in its spherical configuration \citep{Ducci2009}
to the IBIS/ISGRI observations of IGR~J16479$-$4514, 
for which we have a significant number of flares ($N_{\rm fl}=80$).
We will apply this model to the other SFXTs in a future work,
when the number of their flares will be enough.
Our model assumes that a fraction $f$ of the mass
lost by the supergiant wind is in form of clumps 
($f=\dot{M}_{\rm cl}/\dot{M}_{\rm tot}$).
The clumps are assumed to follow a power-law mass distribution 
$p(M_{\rm cl}) \propto M_{\rm cl}^{-\zeta}$
in the mass range $M_{\rm a}-M_{\rm b}$,
a power-law radius distribution
$\dot{N} \propto R_{\rm cl}^{\gamma}$,
and a $\beta$-velocity law (see \citealt{Ducci2009} for details).

We compared the observed and calculated
number of flares, their luminosity distribution,
the X$-$ray luminosity outside flares and
the average time duration of flares.

For each of the 80 flares found, we derived the
peak luminosity in the energy range $1-100$~keV
by means of the spectral parameters found 
by fitting simultaneously the IBIS/ISGRI and JEM-X data 
(see Table \ref{table par. spettri jemx}).
The luminosity ($1-100$~keV) 
outside flares is 
$L_{\rm out-flares} \la 10^{35}$~erg~s$^{-1}$ \citep{Sguera2008}.

To avoid the selection effect given
by the fact that the $5\sigma$ detection threshold
varies with the position of the source
in the field of view,
we have considered only those flares with luminosities greater 
than $L_{\rm lim} \approx 4 \times 10^{36}$~erg~s$^{-1}$,
where we have calculated $L_{\rm lim}$ from the average 
$5\sigma$ detection threshold count$-$rate at an offset angle from the 
pointing position of $\theta = 15^{\circ}$.

\begin{figure}
\begin{center}
\includegraphics[height=6cm]{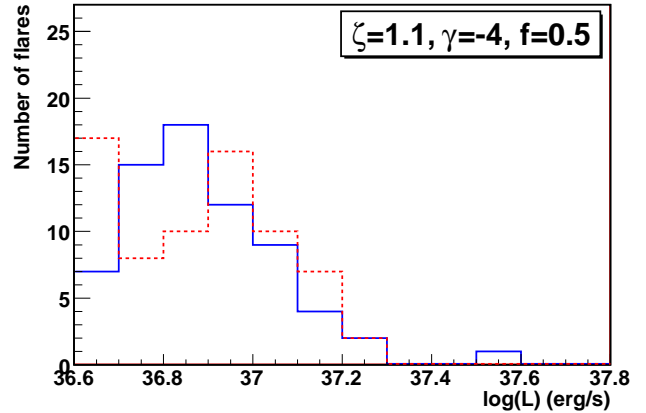}
\end{center}
\caption{Comparison between observed (solid line) and calculated (dashed line) distributions 
of the flare luminosities of IGR~J16479$-$4514.}
\label{istogramma_confronto}
\end{figure}

Figure \ref{istogramma_confronto} shows the 
comparison between the observed and calculated distributions of flare luminosities
which results in the best agreement.
The wind parameters obtained are reported in Table \ref{tab sytstem parameters},
with $\dot{M}=10^{-7}$~M$_\odot$~yr$^{-1}$, 
$v_\infty = 1800$~km~s$^{-1}$, $\beta=1$,
$M_{\rm a}=5 \times 10^{19}$~g, $M_{\rm b}=5 \times 10^{21}$~g, $\zeta=1.1$, $f=0.5$, $\gamma=-4$,
and, since $T_{\rm eff}=34000$~K, the multiplier parameters adopted from \citet{Shimada1994} are:
$k=0.375$, $\alpha=0.522$, $\delta=0.099$ (see \citealt{Ducci2009} for details).
With this set of wind parameters we are able also to reproduce
the observed X$-$ray luminosity outside bright flares and the average flare duration.

\section{Discussion}
\label{sect. discussion}

The wind parameters obtained for IGR~J16479$-$4514 in section \ref{section clumpy wind}
are in agreement with those of the sources previously studied (see \citealt{Ducci2009}; \citealt{Romano2010}),
with the exception of $f$ ($f=0.75$ for the other persistent HMXBs and SFXTs).
Moreover, the mass loss rate found for IGR~J16479$-$4514 ($\dot{M}=10^{-7}$~M$_\odot$~yr$^{-1}$)
is lower if compared to the typical mass loss rate from a O8.5I star,
which is of the order of $\dot{M} \approx 4 \times 10^{-6}$~M$_{\odot}$~yr$^{-1}$
(see Table \ref{tab sytstem parameters}).
This difference could be due to the fact that the
mass loss rates derived from homogeneous-wind
model measurements with the H$\alpha$ method are overestimated 
by a factor 2-10 if the wind is clumpy (see e.g. \citealt{Lepine2008};
\citealt*{Hamann2008}).
Another possibility is that the X$-$ray flaring behaviour in IGR~J16479$-$4514
is not totally due to the accretion of clumps, but other mechanisms
could be at work, like, e.g. the centrifugal inhibition of accretion, 
the formation of transient accretion disks, or the Rayleigh$-$Taylor instability.
The role of these mechanisms in SFXTs will be treated in this section.

\subsection{X$-$ray photoionization} \label{section X-ray photoionization}

\subsubsection{Direct accretion} \label{section Direct accretion}

Here we discuss the effects of X$-$ray photoionization 
on the mass transfer onto a neutron star
in a close binary system, modifying the analytic model
by \citet{Ho1987} and applying it to SFXTs.

In HMXBs, the wind reaching the compact object
is overionized by the X$-$ray photons produced by the compact object \citep{Hatchett1977}.
This high ionization alters the dynamics of the line-driven stellar wind
of the primary: the wind becomes highly ionized and does not interact anymore with 
the UV photons emitted by the primary, which drive the wind.
Hence, the wind velocity and density around the neutron star is
different from what expected neglecting the X$-$ray photoionization.
The accretion onto the compact object is modified accordingly
since it depends onto the wind velocity.

\begin{figure}
\begin{center}
\includegraphics[bb= 0 156 580 516,clip, width=9cm]{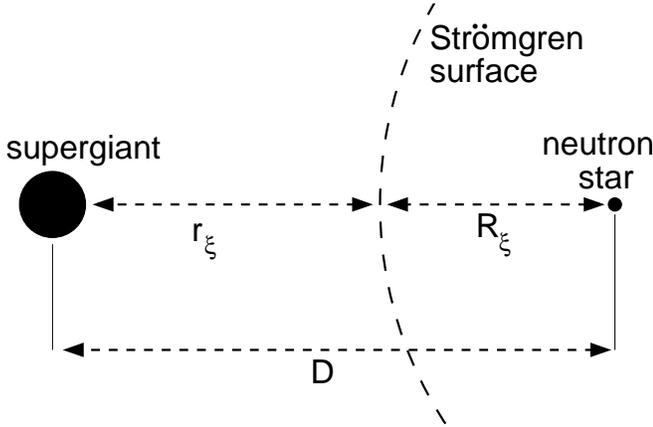}        
\caption{Schematic representation of a Str\"{o}mgren surface produced by a neutron star orbiting
around a supergiant. The labels indicates the distance of this surface
which respect the two stars, accordingly to \citet{Ho1987}.}
\label{Rstromgren}
\end{center}
\end{figure} 

For a source with a X$-$ray luminosity $L_{\rm x}$,
the ionization state $\xi$ of the wind at a radius $R_\xi$ is defined by:
\begin{equation} \label{xilaw}
\xi \equiv \frac{L_{\rm x}}{n(r_\xi)R_\xi^2} 
\end{equation}
where $n(r_\xi)$ is the particle number density at a distance 
$r_\xi= D - R_\xi$ from the supergiant (see Figure \ref{Rstromgren}).
The parameter $\xi$ defined in equation (\ref{xilaw})
determines the thermal and ionic state of the gas,
assuming optically thin gas in thermal balance (see \citealt*{Tarter1969}).

Several attempts have been made to include the effects 
of X$-$ray ionization in the accretion of HMXBs
(e.g. \citealt{Ho1987}; \citealt{MacGregor1982}; \citealt{Blondin1990};
\citealt{Stevens1990}; \citealt{Stevens1991}).
Here we adopt the analytic model developed by \citet{Ho1987}.
They assumed a spherically symmetric wind,
ionized by the X$-$rays from the compact object.
They assumed that the radiation line force is turned off at $\xi_{\rm cr}=10^4$~erg~cm~s$^{-1}$.
Thus, the wind follows the standard 
$\beta$-velocity law (with $\beta=0.5$) up to a 
distance $R_{\rm cr}$ from the compact object such that $L_{\rm x}/[n(R_{\rm cr})R_{\rm cr}^2]=\xi_{\rm cr}$,
i.e. where the wind is sufficiently ionized to become transparent to the UV photons.
Inside this sphere with the neutron star in the centre, 
the radiation force \citep{Kudritzki2000} is turned off
and the wind velocity is assumed constant.
The wind velocity will be lower in the vicinity 
of the compact object (with respect to the non-ionized case), 
leading to an enhancement of the mass accretion rate.
With these assumptions, \citet{Ho1987} developed
a model to describe the accretion of the wind, 
taking into account the feedback effect of the X$-$rays
which ionize the wind, and consequently controls the mass transfer onto the compact object.

We propose the following important changes to this model:
\begin{enumerate}
\item a generic  $\beta$-velocity law,
with $\beta$ not fixed;
\item we consider the orbital velocity of the neutron star in the calculations,
which cannot be neglected in close binary systems such as IGR~J16479$-$4514 
($P_{\rm orb}=3.32$~d, \citealt{Jain2009});
\item we modified the equations developed by \citet{Ho1987}
to take into account the possibility that
the mass loss rate towards the
neutron star is reduced because
of the high ionization state of the wind,
which reduces the radiative acceleration given by
the absorption and re-emission of UV photons
(emitted by the supergiant) in the resonance lines
of ions forming the wind.
Assuming for the wind particles a velocity distribution
centered on $v_{\rm w}(r_\xi)$ (equation \ref{windvelocitylaw}),
a part of them may decelerate enough not to be able to reach the neutron star;
\item a force cutoff value $\xi_{\rm cr}=3 \times 10^2$~erg~cm~s$^{-1}$ \citep{Stevens1991}.
\end{enumerate}

We assumed the standard wind velocity law obtained from the radiation line-driven mechanism
of \citet*{Castor1975}:
\begin{equation} \label{windvelocitylaw}
v_{\rm w}(r_{\xi}) = v_\infty \left ( 1 - \frac{r_{\rm p}}{r_\xi} \right )^\beta
\end{equation}
where $r_\xi$ is the distance from the primary (see Figure \ref{Rstromgren}), and $v_\infty$ is the terminal velocity.

In wind-fed systems, the mass accretion rate 
is defined as the flux of matter passing through 
a circular area with radius $R_{\rm a}$,
and is written as:
\begin{equation} \label{maccr}
  \dot{M}_{\rm accr} = \left [ \frac{\dot{M}}{4 \pi D^2 v_{\rm w}} v_{\rm rel} \right ] \pi R_{\rm a}^2
\end{equation}
where $\dot{M}$ is the wind mass loss rate from the supergiant,
the factor contained in square brackets is the stellar wind flux at a distance $D$
from the primary, $v_{\rm rel} = \sqrt{v_{\rm w}^2 + v_{\rm orb}^2}$ 
is the relative velocity between the wind and the neutron star,
and $R_{\rm a}$ is the accretion radius, defined as:
\begin{equation} \label{Raccr}
R_{\rm a}  = \frac{2 G M_{\rm x}}{(v_{\rm rel}^2 + c_{\rm s}^2)}
\end{equation}
where $M_{\rm x}=1.4$~M$_\odot$ is the mass of the neutron star
and $c_{\rm s}$ is the sound velocity.
Hence, the accretion luminosity, given by equations 
(\ref{windvelocitylaw}, \ref{maccr}, \ref{Raccr}) is:
\begin{equation} \label{Lx eq}
L_{\rm x} = \frac{GM_{\rm x}}{R_{\rm x}} \dot{M}_{\rm accr} = \frac{(GM_{\rm x})^3}{R_{\rm x}} \frac{\dot{M}}{D^2 v_{\rm w}(r_\xi)} \frac{1}{(v_{\rm rel}^2 + c_{\rm s}^2)^{3/2}} \mbox{ .}
\end{equation}

\begin{figure}
\begin{center}
\includegraphics[bb= 100 372 535 700,clip, width=8cm]{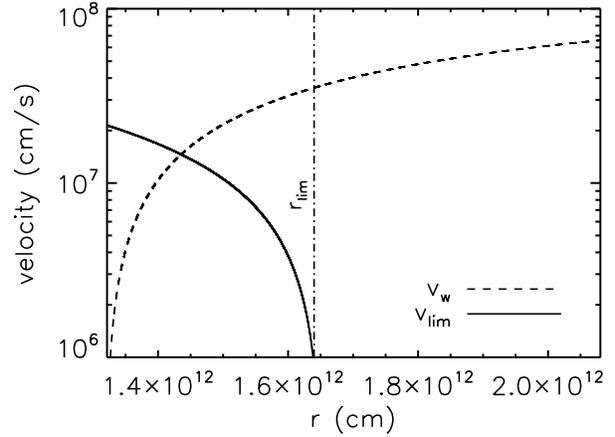}   
\caption{Schematic representation of the wind velocity, 
described by equation (\ref{windvelocitylaw}), and the limit-velocity (equation \ref{solution momentum eq b}).
We assumed $v_\infty=1800$~km~s$^{-1}$, $\beta=1$, $R_{\rm p}=19$~R$_\odot$
and $M_{\rm p}=31$~M$_\odot$, $M_{\rm x}=1.4$~M$_\odot$.}
\label{r-limite.eps}
\end{center}
\end{figure}

The radius $r_{\rm lim}$ where the sum of the radiation force $g_{\rm e}$,
due to scattering of continuum photons by free electrons,
and the gravitational forces of the primary and neutron star 
acting on a parcel of the wind is zero (see Figure \ref{r-limite.eps})
can be derived as follows:
\begin{equation} \label{forze in equilibrio}
\frac{GM_{\rm p}}{r_{\rm lim}^2}(1-\Gamma_{\rm e}) = \frac{GM_{\rm x}}{(D-r_{\rm lim})^2} 
\end{equation}
where $\Gamma_{\rm e}$, defined as
\begin{equation} \label{Gamma formula}
\Gamma_{\rm e} = \frac{\sigma_{\rm e} L_{\rm p}}{4 \pi c G M_{\rm p}} \mbox{ ,}
\end{equation}
is obtained from the formula of
the force due to electron scattering:
\begin{equation} \label{g_e formula}
g_{\rm e} = \frac{\sigma_{\rm e} L_p}{4 \pi c r^2} = \frac{GM_{\rm p}}{r^2} \Gamma_{\rm e} \mbox{ ,}
\end{equation}
where $\sigma_{\rm e}$ is the opacity for electron scattering,
assumed to be equal to $\sim 0.3$~cm$^2$~g$^{-1}$ 
and constant, as suggested by \citet{Lamers1999}.

When the radiation force is turned off,
a parcel of gas is only subject to the gravitational 
force of the two stars and to the force due to electron scattering.
Assuming that this parcel of gas reaches the distance $r_{\rm lim}$ with null velocity,
the initial velocity can be obtained from the momentum equation:
\begin{equation} \label{momentum eq}
v \frac{dv}{dr} = -\frac{G M_{\rm p}}{r^2}(1-\Gamma_{\rm e}) +  \frac{G M_{\rm x}}{(D-r)^2} \mbox{ .}
\end{equation}
For a particle starting at a distance $r$, with velocity $v$,
the solution obtained from equation (\ref{momentum eq}) is:
\begin{equation}
\int_{0}^{v} vdv  =  -\int_{r_{\rm lim}}^r \frac{GM_{\rm p}}{r^{\prime 2}}(1-\Gamma_{\rm e}) dr^{\prime} + \int_{r_{\rm lim}}^r \frac{G M_{\rm x}}{(D-r^{\prime})^2}dr^{\prime} \mbox{.} \label{solution momentum eq a}
\end{equation} 
Thus:
\begin{eqnarray} \label{solution momentum eq b}
v_{\rm lim}(r) & = &  \left [ - 2GM_{\rm p} (1-\Gamma_{\rm e}) \left (\frac{1}{r_{\rm lim}} - \frac{1}{r} \right ) \right. \nonumber \\
     & - & \left. 2GM_{\rm x} \left (\frac{1}{D-r_{\rm lim}} - \frac{1}{D-r} \right ) \right ]^{1/2} \mbox{ .}
\end{eqnarray}

The wind particles have a temperature of $\approx 10^5$~K \citep{Lamers1999}.
Assuming for the particles velocity, at a distance $r$ from the primary,
a gaussian distribution centered on $v_\infty(1-R_{\rm p}/r)^\beta$
with $\sigma_{\rm v} \sim 10^7$~cm~s$^{-1}$,
it is possible to calculate the density probability
to have particles with $v \ga v_{\rm lim}$
at $r=r_{\rm cr}$:
\begin{equation} \label{formula S}
S \equiv \frac{1}{\sqrt{2 \pi} \sigma_{\rm v}} \int_{v_{\rm lim}}^{+\infty} \mbox{e}^{-\frac{(v-v_{\rm w})^2}{2 \sigma_{\rm v}^2}} dv 
\end{equation}
where $v_{\rm w}$ is given by equation (\ref{windvelocitylaw}).

We now propose a new version of the 
\emph{self-consistent steady state equation}
developed by \citet{Ho1987}, improved with the 
considerations described above.
The accretion luminosity $L_{\rm a}$ from equations 
(\ref{windvelocitylaw}), (\ref{Lx eq}), and (\ref{formula S}) is then:
\begin{equation} \label{La}
L_{\rm a}(r_\xi) = \frac{(GM_{\rm x})^3}{R_{\rm x}} \frac{\dot{M}S(r_\xi)}{D^2 v_\infty (1 - R_{\rm p}/r_\xi)^\beta} \frac{1}{(v_{\rm rel}^2 + c_{\rm s}^2)^{3/2}}
\end{equation}
which we call, following the nomenclature of \citet{Ho1987},
the \emph{accretion equation}.
Then we obtain the \emph{feedback equation} from equations
(\ref{xilaw}), (\ref{windvelocitylaw}), (\ref{maccr}), (\ref{Raccr}), (\ref{formula S}):
\begin{equation} \label{Lb}
L_{\rm b}(r_\xi) = \xi_{\rm cr} \frac{\dot{M} S(r_\xi) (D-r_{\xi})^2}{4 \pi r_\xi^2 \mu m_{\rm p} v_\infty (1 -R_{\rm p}/r_\xi)^\beta} 
\end{equation}
where $\mu$ is the mean atomic weight.
The steady state solution is given by:
$L_{\rm a}(r_\xi) = L_{\rm b}(r_\xi)$,
thus, by equating equations (\ref{La}) and (\ref{Lb}),
it is possible to obtain the X$-$ray luminosity $L_{\rm x}$
and the corresponding Str\"{o}mgren radius $r_{\rm cr}$.

\begin{table}
\begin{center}
\caption{System Parameters for IGR~J16479$-$4514 and IGR~J17544$-$2619.}
\label{tab sytstem parameters}
\begin{tabular}{lcc}
\hline
Parameters                    & \multicolumn{2}{c}{Sources} \\
                              &                IGR~J16479$-$4514               &            IGR~J17544$-$2619             \\
\hline
Spectral type                 &              O8.5I $^{\mathrm{a}}$             &         O9I $^{\mathrm{a}}$              \\ 
$P_{\rm orb}$ (d)             &       $3.3194 \pm 0.0010^{\mathrm{b}}$     &     $4.926 \pm 0.001^{\mathrm{g}}$   \\
$M_{\rm p}$ ($M_\odot$)        &             $31$~$^\mathrm{c}$                 &          $25-28$~$^\mathrm{g}$          \\
$R_{\rm p}$ ($R_\odot$)        &             $19$~$^\mathrm{c}$                 &      $<23$ if $e=0$~$^\mathrm{g}$        \\
                               &                                               &$\sim 12.7$ if $e \sim 0.4$~$^\mathrm{g}$ \\
$L_{\rm p}$ ($L_\odot$)        &     $\approx 5\times 10^5$~$^\mathrm{d}$       & $\approx 4.6\times 10^5$~$^\mathrm{d}$   \\ 
$\dot{M}$ ($M_\odot$~yr$^{-1}$) &   $\sim 4 \times 10^{-6}$~$^\mathrm{e}$       & $\sim 2.4 \times 10^{-6}$~$^\mathrm{e}$   \\
$M_{\rm x}$ ($M_\odot$)         &           $1.4$~$^\mathrm{f}$                 &           $1.4$~$^\mathrm{f}$            \\
$v_\infty$ (km~s$^{-1}$)        &                  $1800$                      &                $1800$                   \\
$\beta$                       &                     $1$                      &                    $1$                   \\
\hline
\end{tabular}
\end{center}
  \begin{list}{}{}
  \item[$^{\mathrm{a}}$]{\citet{Rahoui2008}}
  \item[$^{\mathrm{b}}$]{\citet{Jain2009}}
  \item[$^{\mathrm{c}}$]{obtained assuming no Roche-Lobe overflow (see section \ref{section clumpy wind})}
  \item[$^{\mathrm{d}}$]{\citet{Vacca1996}; \citet{Martins2005}}
  \item[$^{\mathrm{e}}$]{Calculated with the Vink formula \citep*{Vink2000}}
  \item[$^{\mathrm{f}}$]{Assumed}
  \item[$^{\mathrm{g}}$]{\citet{Clark2009}}
  \end{list}
\end{table}

We assume for the force cutoff value $\xi_{\rm cr}=3 \times 10^2$~erg~cm~s$^{-1}$,
which is two orders of magnitude 
smaller than the value considered by \citet{Ho1987}, 
but in agreement with the calculations of \citet{Stevens1991}, 
who found that for this value the wind material 
is already basically completely ionized.

We calculate the X$-$ray luminosity and $r_{\rm cr}$
of IGR~J16479$-$4514 from the steady state solution obtained from
equations (\ref{La}) and (\ref{Lb}),
assuming the parameters reported in Table \ref{tab sytstem parameters}.
We obtain an expected luminosity  of $L_{\rm x} \approx 10^{37}$~erg~s$^{-1}$,
reached by the neutron star when $r_{\rm cr}=1.33 \times 10^{12}$~cm.
This luminosity level is in agreement with the peak flare luminosities observed.
However, IGR~J16479$-$4514 has an out-of-flare luminosity 
of the order of $10^{34}-10^{35}$~erg~s$^{-1}$ \citep{Sidoli2008}.
This luminosity level can be obtained possibly invoking 
the presence of the centrifugal inhibition of accretion.

For IGR~J17544$-$2619 we assumed the system parameters 
suggested by \citet{Clark2009}, with eccentricity $e=0.4$.
We found that the X$-$ray photoionization acts to increase the difference in luminosity
when the neutron star is at periastron ($L_{\rm x} \approx 10^{37}$~erg~s$^{-1}$) 
and apastron ($L_{\rm x} \approx 10^{35}$~erg~s$^{-1}$, see Table \ref{tab model results2}).
In fact, at periastron the wind is highly ionized by the X$-$ray source
($r_{\rm cr} \simeq 0.9 \times 10^{12}$~cm), leading to a low
wind velocity in the vicinity of the neutron star ($v_{\rm rel} \simeq 230$~km~s$^{-1}$)
and accordingly to a higher X-ray luminosity (since $L_{\rm x} \propto v_{\rm rel}^{-3}$).
At apastron the ionization is lower ($r_{\rm cr} \simeq 3.3 \times 10^{12}$~cm),
therefore its effect on the wind velocity is small ($v_{\rm rel} \simeq 1400$~km~s$^{-1}$).

\begin{table}
\begin{center}
\caption{Results for IGR~J17544$-$2619.}
\label{tab model results2}
\begin{tabular}{lcc}
\hline
                                &       Periastron          &        Apastron             \\
\hline
$\dot{M}$ (M$_\odot$~yr$^{-1}$)  &     $2.4 \times 10^{-6}$   &    $2.4 \times 10^{-6}$       \\
$r_{\rm cr}$ (cm)                & $\sim 0.9 \times 10^{12}$  & $\sim 3.3 \times 10^{12}$     \\
$L_{\rm x}$  (erg~s$^{-1}$)      &       $\ga 10^{37}$        &   $\sim 2 \times 10^{35}$      \\
$v_{\rm rel}$ (km~s$^{-1}$)       &       $\sim 230$          &      $\sim 1400$              \\
$R_{\rm m}$  (cm)                &        $10^{10}$           &   $4.5 \times 10^{10}$        \\
\hline
\end{tabular}
\end{center}
\end{table}

\subsubsection{Centrifugal inhibition of accretion}

In this section we apply for the first time the
centrifugal inhibition (c.i.) of accretion to the modified model of \citet{Ho1987}
described in section \ref{section Direct accretion}.

The magnetospheric radius $R_{\rm m}$ is defined as the radius where 
the magnetic field pressure $B^2(R_{\rm m})/8\pi$ 
equals the ram pressure of the accreting plasma $\rho(R_{\rm m}) v^2(R_{\rm m})$.
We use the definition obtained by \citet{Davidson1973}:
\begin{equation} \label{Rm Davidson}
R_{\rm m}(r_\xi) = \left [ \frac{B_{\rm 0}^2 R_{\rm x}^6}{4 \dot{M}_{\rm accr} (G M_{\rm x})^{1/2}} \right ]^{2/7}
\end{equation}
where $B_{\rm 0}$ is the surface magnetic field.
When $R_{\rm m} \ga R_{\rm co}$ (where $R_{\rm co}$ is the corotation radius), 
the accretion flow is halted at the magnetospheric boundary,
which behaves like a closed barrier \citep{Illarionov1975}.
The expected X$-$ray luminosity in this regime is:
\begin{equation}
L_{\rm m} = \frac{GM}{R_{\rm m}} \dot{M}_{\rm accr} \mbox{ .}
\end{equation}
The position of the magnetospheric radius depends also on
the effect of the X$-$ray ionization on the accreting matter.
Here we obtain the formula which gives the steady position
of the magnetospheric radius and the related position of the Str\"{o}mgren
radius for which the accretion luminosity is equal to the feedback luminosity.
The accretion equation (\ref{La}) can be written as follows:
\begin{equation} \label{La2}
L_{\rm a, c.i.}(r_\xi) = \frac{(GM_{\rm x})^3}{R_{\rm m}} \frac{\dot{M}S(r_\xi)}{D^2 v_\infty (1 - R_{\rm p}/r_\xi)^\beta} \frac{1}{(v_{\rm rel}^2 + c_{\rm s}^2)^{3/2}} \mbox{.}
\end{equation}
We obtain $R_{\rm m}$ as a function of $r_{\rm cr}$
by equating $L_{\rm b}$ of the feedback equation (\ref{Lb})
to the $L_{\rm a, c.i.}$ of the accretion equation (\ref{La2}):
\begin{equation} \label{Rm ab}
R_{\rm m, c.i.}(r_\xi) = \frac{(G M_{\rm x})^3 4 \pi \mu m_{\rm p}}{(v_{\rm rel}^2 +c_{\rm s}^2)^{3/2} \xi_{\rm cr} (D-r_\xi)^2} \mbox{ .}
\end{equation}
Hence, if the centrifugal inhibition of accretion is at work in an X$-$ray binary system
where the X$-$ray ionization cannot be neglected, 
the steady values for $R_{\rm m}$ and $r_{\rm cr}$ can be derived 
by equating equations (\ref{Rm Davidson}) and (\ref{Rm ab}):
\begin{equation} \label{condizione Rm stabile}
R_{\rm m}(r_\xi) = R_{\rm m, c.i.}(r_\xi) \mbox{ .}
\end{equation}

In Table \ref{tab model results} we report the steady state solutions obtained 
for IGR~J16479$-$4514 (assuming the system parameters reported 
in Table \ref{tab sytstem parameters}) in the case of direct accretion 
and in the case of centrifugal inhibition of accretion.
Assuming a transition from direct accretion to 
the centrifugal inhibition of accretion for a mass-loss rate of 
$\dot{M} = 4 \times 10^{-6}$~M$_\odot$~yr$^{-1}$,
we obtain steady state solutions in agreement with observations for a magnetic field
of $B_{\rm 0} \approx 10^{12}$~Gauss and a spin period of $\approx 1$~s.
When the neutron star enters the state of centrifugal inhibition of accretion,
the X$-$ray luminosity is reduced, thus also the photoionization is lower;
this leads to a shift of the magnetospheric radius to a higher value.
The opposite transition (from centrifugal inhibition to direct accretion)
needs $\dot{M}$ greater than  $4 \times 10^{-6}$~M$_\odot$~yr$^{-1}$
(e.g. the accretion of a dense clump).
This behaviour allows longer durations for the low-luminosity state.
When $R_{\rm m} \approx R_{\rm co}$, the oblate spheroid shape 
of the magnetospheric boundary (see \citealt*{Jetzer1998}) 
results in the intermediate luminosity 
state observed ($\sim 10^{35}$~erg~s$^{-1}$)
because simultaneously $R_{\rm m} < R_{\rm co}$ in the magnetic polar region
and $R_{\rm m} > R_{\rm co}$ in the magnetic equatorial region
(\citealt{Campana2001}; \citealt*{Perna2006}).

\begin{table}
\begin{center}
\caption{Model results for IGR~J16479$-$4514.
We calculate the magnetospheric radius $R_{\rm m}$ with equation 
(\ref{Rm Davidson}), assuming a magnetic field of $B_{\rm 0}=10^{12}$~Gauss.
For IGR~J16479$-$4514, the corotation radius is equal 
to the magnetospheric radius when 
$\dot{M} = 4 \times 10^{-6}$~M$_\odot$~yr$^{-1}$ 
for a spin period of the neutron star of $\sim 1$~s.}
\label{tab model results}
\begin{tabular}{lcc}
\hline
                                &       Direct accretion    & Centrifugal inhibition                 \\
                                & ($R_{\rm m} < R_{\rm co}$): & of accretion ($R_{\rm m} > R_{\rm co}$): \\
\hline
$\dot{M}$ (M$_\odot$~yr$^{-1}$)  &  $\ga 4 \times 10^{-6}$    &    $4 \times 10^{-6}$                  \\
$r_{\rm cr}$ (cm)                & $\sim 1.33 \times 10^{12}$ & $\sim 2 \times 10^{12}$                \\
$L_{\rm x}$  (erg~s$^{-1}$)      &       $\ga 10^{37}$        &  $\ga 8 \times 10^{34}$                \\
$v_{\rm rel}$ (km~s$^{-1}$)       &       $\sim 460$          &      $\sim 760$                        \\
$R_{\rm m}$  (cm)                &    $1.6 \times 10^8$      &   $1.7 \times 10^8$                    \\
\hline
\end{tabular}
\end{center}
\end{table}

\subsection{Formation of an accretion disk} \label{section formation accr disk}

\begin{figure}
\begin{center}
\includegraphics[width=9cm, bb=100 360 558 720,clip]{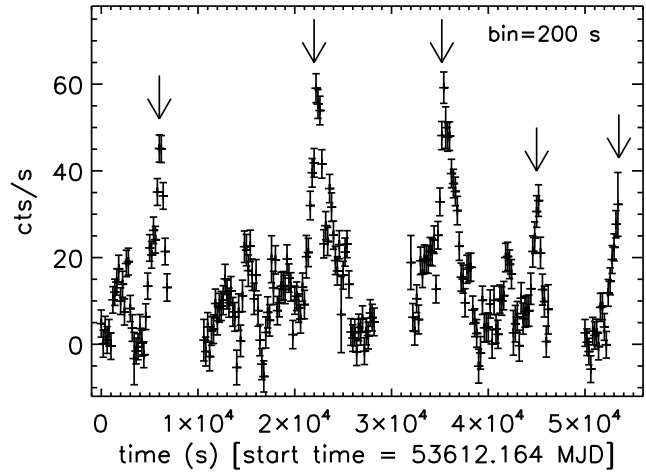}
\end{center}
\caption{Lightcurve of the observation of XTE~J1739$-$302, 
in the energy range $18-60$~keV (IBIS/ISGRI).
Arrows indicates the peaks of luminosity.}
\label{lcrsxtej1739}
\end{figure}

In the previous section we discussed the effect of X$-$ray photoionization 
in a close X$-$ray binary system like IGR~J16479$-$4514,
where a compact object accretes matter from a wind 
with a low relative velocity $v_{\rm rel} \approx 460$~km~s$^{-1}$.
In section \ref{section Direct accretion} we have excluded for this system 
the possibility of the formation of an accretion disk 
due to Roche Lobe overflow.
However, it is also possible that an accretion disk forms from the capture of 
angular momentum from a slow wind (see \citealt{Illarionov1975};
\citealt{Shapiro1976}; \citealt{Wang1981}). 
The existence of a disk in IGR~J16479$-$4514 requires that the relative 
velocity between the neutron star
and the wind is lower than $\simeq 500$~km~s$^{-1}$ 
(see formula [31] in \citealt{Wang1981}),
therefore the presence of a disk in IGR~J16479$-$4514 cannot be ruled out.
The formation of a transient disk from the mass and angular momentum capture
from an asymmetric stellar wind can lead to a flaring activity, as proposed
by \citet{Taam1988} to explain the recurrent flares observed in EXO~2030+375 \citep{Parmar1989}.
Recently, the formation of transient disks has been proposed by
\citet{Kreykenbohm2008} to explain part of the flaring behaviour of the
persistent HMXB Vela X$-$1.
In the model of \citet{Taam1988}, the density and velocity inhomogeneities 
in the wind lead to an instability in the accretion flow.
Because of this instability, the interaction of the accretion flow
with the shock fronts of the accretion wake leads to the formation
and dissipation of transient accretion disks (see e.g. \citealt{Edgar2004} and references therein
for a recent review).
In their model, \citet{Taam1988} gave an approximative formula for the 
time scale for the duration of flares:
\begin{equation} \label{tau taam}
\tau \sim \frac{6GM_{\rm x}}{v_{\rm rel}^3} \mbox{ .}
\end{equation}
Assuming relative velocities reported in Table \ref{tab model results}
(in the case of direct accretion), we found for IGR~J16479$-$4514 
a time scale for the duration of flares
of $\sim 10^3-10^4$~s, in agreement with the observed flare durations.
The flares produced by the formation of transient accretion disks
can be distinguished from those produced by the simple clump accretion 
by measuring the time derivative of the pulse period,
which, in the first case, is expected to change sign for each flare (see \citealt{Taam1988}).

The SFXT XTE~J1739$-$302 has shown a quasi-periodic flaring behaviour on 53612.164 MJD
(see Figure \ref{lcrsxtej1739}), which could be ascribed 
to the mechanism of formation and dissipation of transient accretion disks 
described above.
Adopting equation (\ref{tau taam}),
we obtain, from the measure of the flares durations ($\tau \sim 5000$~s),
a relative velocity of $v_{\rm rel} \approx 600$~km~s$^{-1}$.
This value is in agreement with the $v_{\rm rel}$ expected
from an X$-$ray binary system with a small orbital period ($\approx 3-7$~d),
or with a higher orbital period and large eccentricity, 
where the effect of X$-$ray photoionization reduces significantly the wind velocity.
This supports the hypothesis of \citet{Smith2006} that the fast outbursts
of XTE~J1739$-$302 could be due to an instability of an accretion disk.

\subsection{Accretion via Rayleigh$-$Taylor instability}
\label{rti}

\begin{figure}
\begin{center}
\includegraphics[width=9cm, bb=90 360 558 720,clip]{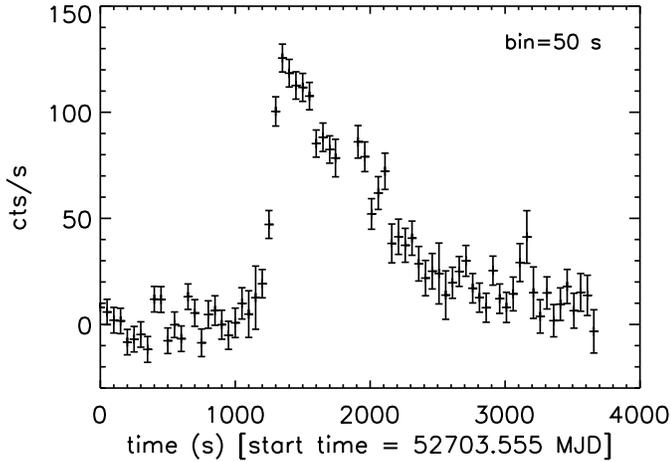}
\end{center}
\caption{Lightcurve of an observation of IGR~J16479$-$4514, 
in the energy range $18-60$~keV (IBIS/ISGRI).}
\label{lcrsigrj16479}
\end{figure}

The shape of the flare of IGR~J16479$-$4514 
we observed with \emph{INTEGRAL} and 
reported in Figure \ref{lcrsigrj16479},
i.e. fast rise and exponential decay, could be explained with the 
magnetospheric instability mechanism proposed by \citet{Lamb1977}
to explain type II X$-$ray bursts.
All equations in this section are taken from these authors.
In the framework of a spherically symmetric accretion flow onto a neutron star,
and under particular conditions of X$-$ray luminosity and of the temperature
of the accreting matter, the magnetospheric surface behaves as a gate
which controls the flow towards the stellar surface.
When the gate is closed, a magnetospheric cavity is formed,
and a reservoir of matter is accumulated on the top of the neutron star magnetosphere,
leading to low X$-$ray luminosities.
When this matter has cooled enough (because of electron$-$ion bremsstrahlung,
which is the dominant cooling mechanism when the gate is closed),
the plasma enters the magnetosphere via Rayleigh$-$Taylor instability,
and the accretion onto the neutron star leads to a  flaring behaviour.
\citet{Lamb1977} found a critical luminosity
\begin{equation} \label{L crit Lamb}
L_{\rm crit} = 2 \times 10^{36} |1 - T_{\rm r}/T_{\rm c}|^{7/8} \mu_{30}^{1/4} \left ( \frac{M}{M_\odot}\right )^{1/2} R_6^{-1/8} \mbox{erg s}^{-1}
\end{equation}
where $T_{\rm r}$ is the temperature of radiation,
$T_{\rm c} \approx 10^9$~K, $\mu_{30}$ is the stellar magnetic moment
in units of $10^{30}$~Gauss~cm$^3$, $M$ is the mass of the neutron star
and $R_6$ is the radius of the neutron star in units of $10^6$~cm.
If the flare luminosity is greater than the critical luminosity (equation \ref{L crit Lamb}),
the Compton cooling dominates the bremsstrahlung cooling at $R_{\rm m}$,
thus the magnetosphere gate is open for longer time and the flare is prolonged \citep{Lamb1977}.

The flare reported in Figure \ref{lcrsigrj16479} has a peak luminosity of
$\approx 3 \times 10^{37}$~erg~s$^{-1}$, which is greater than the critical luminosity
$L_{\rm crit} \approx 2.4 \times 10^{36}$~erg~s$^{-1}$ (if $T_{\rm r} < 10^9$~K).

Assuming $T_{\rm r} \approx 10^8$~K, we found that, because of Compton processes,
beyond the radius $r_{\rm c} \approx 2.2 \times 10^{10}$~cm at which the free-fall
temperature equals the temperature of radiation, 
the X$-$rays emitted by the neutron star heat the 
plasma\footnote{The material accreting onto a neutron star
is decelerated by the magnetic field, hence it is shock-heated to
a temperature $T_{\rm s} \approx 3/16 T_{\rm ff}(R_{\rm m})$,
where $T_{\rm ff}(r) \propto 1/r$ is the proton free-fall temperature.
Since the temperature of radiation is roughly constant for every $r$,
below a radius $r_{\rm c}$ the gas is cooled
by Compton interaction ($T_{\rm r} < T_{\rm gas}$);
outside $r_{\rm c}$ we have $T_{\rm r} > T_{\rm gas}$,
thus the gas is heated by Compton processes.}
(see formula [31] in \citealt{Lamb1977}).

If the heating time scale 
is less than the flow time scale\footnote{The condition for which the heating time
scale is less than the flow time scale is expressed by:
\begin{equation} \label{cond. lamb1}
L_{\rm x} \gg L_{\rm choke} = 1.2 \times 10^{37} \left ( \frac{T_{\rm r}}{10^8 K}\right )^{-1/2} \left ( \frac{M}{M_{\odot}} \right ) \mbox{ erg~s}^{-1} \mbox{ .}
\end{equation}
This condition is respected if $T_{\rm r}=10^8$~K, 
$M=1.4 M_\odot$, $L_{\rm x} =3 \times 10^{37}$~erg~s$^{-1}$.
}, 
the flow is choked at $R_{\rm m}$, and we expect for the flare
a time duration given by:
\begin{equation} \label{cond lamb2}
\Delta t_{\rm max} \approx 125 \left ( \frac{T_{\rm r}}{10^8 K} \right )^{-3/2} \left ( \frac{M}{M_{\odot}} \right ) \mbox{ s}
\end{equation}
which results into $\Delta t_{\rm max} \approx 175$~s assuming 
$T_{\rm r}=10^8$~K and $M=1.4 M_\odot$.
This duration is not in agreement with that observed in \mbox{Figure} \ref{lcrsigrj16479}.
However, if we assume $T_{\rm r} \approx 2 \times 10^7$~K, 
the heating time scale is less than the flow
timescale at distances $\gg r_{\rm c}$.
The condition (\ref{cond. lamb1}) is no longer valid 
($L_{\rm x} \la L_{\rm choke} \approx 4 \times 10^{37}$~erg~s$^{-1}$),
thus the flow is choked at a radius $r_{\rm c}' > r_{\rm c}$.
In this case the expected burst duration is given by:
\begin{equation} \label{cond lamb3}
\Delta t_{\rm max}' \approx 210 L_{\rm 37}^{-3} \left ( \frac{T_{\rm r}}{10^8 K}\right )^{-3} \left ( \frac{M}{M_{\odot}} \right )^4 \mbox{ s} \approx 4900 \mbox{ s} 
\end{equation}
(assuming $T_{\rm r}=2 \times 10^7$~K, $M=1.4 M_\odot$ and $L_{\rm x} =3 \times 10^{37}$~erg~s$^{-1}$).
The assumption $T_{\rm r}=2 \times 10^7$~K is in agreement with
the temperature of radiation observed in some SFXTs 
(see e.g. \citealt{Sidoli2007}).
In the framework described by \citet{Lamb1977},
the exponential decay in this kind of flares
is described by the law $L(t) \propto t^{1-2\alpha /3}$,
with $\alpha \ga 3/2$.
From the fit of the exponential decay of the flare observed by \emph{INTEGRAL} (Figure \ref{lcrsigrj16479})
we found $\alpha \approx 1.55$,
thus in agreement with the value predicted by \citet{Lamb1977}.

We point out that if $L_{\rm x} > L_{\rm crit}$,
$T_{\rm r} < T_{\rm c}$ at $R_{\rm m}$,
and if the accretion radius is lower than $r_{\rm c}$,
the X$-$ray source is persistent with high luminosity (see \citealt{Elsner1977}).
In our case $R_{\rm a} \approx 1.8 \times 10^{11}$~cm
and $r_{\rm c} \approx 2.2 \times 10^{10}$~cm (if $T_{\rm r} = 10^8$~K)
or $r_{\rm c} \approx 1.1 \times 10^{11}$~cm (if $T_{\rm r} = 2 \times 10^7$~K).
Thus, what makes IGR~J16479$-$4514 (and probably other SFXTs) intermittent 
with respect to the other persistent HMXBs could be the X$-$ray
photoionization which reduces the wind velocity,
increasing the accretion radius
with respect to $r_{\rm c}$.

\section{Conclusions} \label{Conclusions}

Up to now three accretion mechanisms have been proposed 
to explain the X$-$ray flares of SFXTs.

Here we propose other mechanisms to explain the observed X$-$ray behaviour of
IGR~J16479$-$4514, IGR~J17544$-$2619 and XTE~J1739$-$302: 
the effect of X$-$ray photoionization onto accretion
(both in direct accretion and centrifugal inhibition of accretion)
in the framework of the \citet{Ho1987} model;
the accretion disk instability of \citet{Taam1988}, and the Rayleigh-Taylor instability of \citet{Lamb1977}.

We have shown that the X$-$ray photoionization reduces the mass loss rate
and the wind velocity along the trajectory compact object$-$OB supergiant
with respect to the undisturbed case (section \ref{section X-ray photoionization}).
Their simultaneous reduction leads to X$-$ray 
luminosities in agreement with 
those observed during the flaring activity; moreover, a lower $v_{\rm w}$ allows 
the formation of transient accretion disks from the capture of angular momentum,
able to reproduce some kind of quasi$-$periodic 
recurrent flares observed in SFXTs.

We found in our analysis of \emph{INTEGRAL} data 
that some flares show a peculiar shape, 
characterized by a sharp rise and an exponential decay (Figure \ref{lcrsigrj16479}).
We found that this shape is well explained assuming
the model of \citet{Lamb1977}, based on the Rayleigh-Taylor instability,
although this model was proposed to try to explain type II bursts.

Therefore, we conclude that in SFXTs with large orbital periods 
($P_{\rm orb} \ga 15$~d; e.g.: IGR~J18483$-$0311, SAX~J1818.6$-$1703,
IGR~J11215-5952) the effects of X$-$ray photoionization onto accretion mechanism
can be neglected.
In SFXTs with smaller orbital periods ($P_{\rm orb} \la 15$~d; e.g.: IGR~J16479$-$4514, IGR~J17544$-$2619)
we expect that part of the X$-$ray variability observed is due to the 
X$-$ray photoionization and consequently to the accretion disk instability.
We are still not able to establish which mechanism dominates in SFXTs with smaller orbital periods.
Moreover, we point out that different accretion mechanisms could be at work
in a single SFXT, in the case of orbits with high eccentricities.

\section*{Acknowledgments}

Based on observations with \textit{INTEGRAL}, an ESA project
with instruments and science data centre funded by ESA member states
(especially the PI countries: Denmark, France, Germany, Italy,
Spain, and Switzerland), Czech Republic and Poland, and with the
participation of Russia and the USA.\\
AP acknowledges the Italian Space Agency financial support via contract
I/008/07/0.

\bibliographystyle{mn2e}
\bibliography{lducci_sfxts.bib}

\begin{thebibliography}{}

\bibitem[\protect\citeauthoryear{{Blondin}, {Kallman}, {Fryxell} \&
  {Taam}}{{Blondin} et~al.}{1990}]{Blondin1990}
{Blondin} J.~M.,  {Kallman} T.~R.,  {Fryxell} B.~A.,    {Taam} R.~E.,  1990,
  \apj, 356, 591

\bibitem[\protect\citeauthoryear{{Bozzo}, {Falanga} \& {Stella}}{{Bozzo}
  et~al.}{2008a}]{Bozzo2008a}
{Bozzo} E.,  {Falanga} M.,    {Stella} L.,  2008a, \apj, 683, 1031

\bibitem[\protect\citeauthoryear{{Bozzo}, {Stella}, {Israel}, {Falanga} \&
  {Campana}}{{Bozzo} et~al.}{2008b}]{Bozzo2008b}
{Bozzo} E.,  {Stella} L.,  {Israel} G.,  {Falanga} M.,    {Campana} S.,  2008b,
  \mnras, 391, L108

\bibitem[\protect\citeauthoryear{{Bozzo}, {Stella}, {Ferrigno}, {Giunta},
  {Falanga}, {Campana}, {Israel} \& {Leyder}}{{Bozzo} et~al.}{2010}]{Bozzo2010}
{Bozzo} E.,  {Stella} L.,  {Ferrigno} C.,  {Giunta} A.,  {Falanga} M.,
  {Campana} S.,  {Israel} G.,    {Leyder} J.~C.,  2010, ArXiv e-prints


\bibitem[\protect\citeauthoryear{{Campana}, {Gastaldello}, {Stella}, {Israel},
  {Colpi}, {Pizzolato}, {Orlandini} \& {Dal Fiume}}{{Campana}
  et~al.}{2001}]{Campana2001}
{Campana} S.,  {Gastaldello} F.,  {Stella} L.,  {Israel} G.~L.,  {Colpi} M.,
  {Pizzolato} F.,  {Orlandini} M.,    {Dal Fiume} D.,  2001, \apj, 561, 924

\bibitem[\protect\citeauthoryear{{Castor}, {Abbott} \& {Klein}}{{Castor}
  et~al.}{1975}]{Castor1975}
{Castor} J.~I.,  {Abbott} D.~C.,    {Klein} R.~I.,  1975, \apj, 195, 157

\bibitem[\protect\citeauthoryear{{Clark}, {Hill}, {Bird}, {McBride}, {Scaringi}
  \& {Dean}}{{Clark} et~al.}{2009}]{Clark2009}
{Clark} D.~J.,  {Hill} A.~B.,  {Bird} A.~J.,  {McBride} V.~A.,  {Scaringi} S.,
    {Dean} A.~J.,  2009, \mnras, 399, L113

\bibitem[\protect\citeauthoryear{{Davidson} \& {Ostriker}}{{Davidson} \&
  {Ostriker}}{1973}]{Davidson1973}
{Davidson} K.,  {Ostriker} J.~P.,  1973, \apj, 179, 585

\bibitem[\protect\citeauthoryear{{Ducci}, {Sidoli}, {Mereghetti}, {Paizis} \&
  {Romano}}{{Ducci} et~al.}{2009}]{Ducci2009}
{Ducci} L.,  {Sidoli} L.,  {Mereghetti} S.,  {Paizis} A.,    {Romano} P.,
  2009, \mnras, 398, 2152

\bibitem[\protect\citeauthoryear{{Edgar}}{{Edgar}}{2004}]{Edgar2004}
{Edgar} R.,  2004, New Astronomy Review, 48, 843

\bibitem[\protect\citeauthoryear{{Eggleton}}{{Eggleton}}{1983}]{Eggleton1983}
{Eggleton} P.~P.,  1983, \apj, 268, 368

\bibitem[\protect\citeauthoryear{{Elsner} \& {Lamb}}{{Elsner} \&
  {Lamb}}{1977}]{Elsner1977}
{Elsner} R.~F.,  {Lamb} F.~K.,  1977, \apj, 215, 897

\bibitem[\protect\citeauthoryear{{Goldwurm}, {David}, {Foschini}, {Gros},
  {Laurent}, {Sauvageon}, {Bird}, {Lerusse} \& {Produit}}{{Goldwurm}
  et~al.}{2003}]{Goldwurm2003}
{Goldwurm} A. et al.,    2003, \aap,
  411, L223

\bibitem[\protect\citeauthoryear{{Hamann}, {Feldmeier} \& {Oskinova}}{{Hamann}
  et~al.}{2008}]{Hamann2008}
{Hamann} W.-R.,  {Feldmeier} A.,    {Oskinova} L.~M.,  eds, 2008, {Clumping in
  hot-star winds, University of Potsdam, Potsdam, p.11}

\bibitem[\protect\citeauthoryear{{Hatchett} \& {McCray}}{{Hatchett} \&
  {McCray}}{1977}]{Hatchett1977}
{Hatchett} S.,  {McCray} R.,  1977, \apj, 211, 552

\bibitem[\protect\citeauthoryear{{Ho} \& {Arons}}{{Ho} \&
  {Arons}}{1987}]{Ho1987}
{Ho} C.,  {Arons} J.,  1987, \apj, 316, 283

\bibitem[\protect\citeauthoryear{{Illarionov} \& {Sunyaev}}{{Illarionov} \&
  {Sunyaev}}{1975}]{Illarionov1975}
{Illarionov} A.~F.,  {Sunyaev} R.~A.,  1975, \aap, 39, 185

\bibitem[\protect\citeauthoryear{{in't Zand}}{{in't Zand}}{2005}]{intZand2005}
{in't Zand} J.~J.~M.,  2005, \aap, 441, L1

\bibitem[\protect\citeauthoryear{{Jain}, {Paul} \& {Dutta}}{{Jain}
  et~al.}{2009}]{Jain2009}
{Jain} C.,  {Paul} B.,    {Dutta} A.,  2009, \mnras, 397, L11

\bibitem[\protect\citeauthoryear{{Jetzer}, {Strassle} \& {Straumann}}{{Jetzer}
  et~al.}{1998}]{Jetzer1998}
{Jetzer} P.,  {Strassle} M.,    {Straumann} N.,  1998, New Astronomy, 3, 619

\bibitem[\protect\citeauthoryear{{Kreykenbohm}, {Wilms}, {Kretschmar},
  {Torrej{\'o}n}, {Pottschmidt}, {Hanke}, {Santangelo}, {Ferrigno} \&
  {Staubert}}{{Kreykenbohm} et~al.}{2008}]{Kreykenbohm2008}
{Kreykenbohm} I. et al.,  2008, \aap, 492, 511

\bibitem[\protect\citeauthoryear{{Kudritzki} \& {Puls}}{{Kudritzki} \&
  {Puls}}{2000}]{Kudritzki2000}
{Kudritzki} R.,  {Puls} J.,  2000, \araa, 38, 613

\bibitem[\protect\citeauthoryear{{Labanti}, {Di Cocco}, {Ferro}, {Gianotti},
  {Mauri}, {Rossi}, {Stephen}, {Traci} \& {Trifoglio}}{{Labanti}
  et~al.}{2003}]{Labanti2003}
{Labanti} C. et al.,  2003, \aap,
  411, L149

\bibitem[\protect\citeauthoryear{{Lamb}, {Fabian}, {Pringle} \& {Lamb}}{{Lamb}
  et~al.}{1977}]{Lamb1977}
{Lamb} F.~K.,  {Fabian} A.~C.,  {Pringle} J.~E.,    {Lamb} D.~Q.,  1977, \apj,
  217, 197

\bibitem[\protect\citeauthoryear{{Lamers} \& {Cassinelli}}{{Lamers} \&
  {Cassinelli}}{1999}]{Lamers1999}
{Lamers} H.~J.~G.~L.~M.,  {Cassinelli} J.~P.,  1999, in Henny J.~G.~L.~M.~Lamers and Joseph
  P.~Cassinelli, eds, {Introduction to Stellar Winds}. Cambridge Univ. Press, Cambridge, 
  p.~452. (ISBN 0521593980)

\bibitem[\protect\citeauthoryear{{Lebrun}, {Leray}, {Lavocat}, {Cr{\'e}tolle},
  {Arqu{\`e}s}, {Blondel}, {Bonnin}, {Bou{\`e}re}, {Cara}, {Chaleil}, {Daly} \&
  {Desages}}{{Lebrun} et~al.}{2003}]{Lebrun2003}
{Lebrun} F. et al.,  2003, \aap, 411, L141

\bibitem[\protect\citeauthoryear{{L{\'e}pine} \& {Moffat}}{{L{\'e}pine} \&
  {Moffat}}{2008}]{Lepine2008}
{L{\'e}pine} S.,  {Moffat} A.~F.~J.,  2008, \aj, 136, 548

\bibitem[\protect\citeauthoryear{{Leyder}, {Walter}, {Lazos}, {Masetti} \&
  {Produit}}{{Leyder} et~al.}{2007}]{Leyder2007}
{Leyder} J.,  {Walter} R.,  {Lazos} M.,  {Masetti} N.,    {Produit} N.,  2007,
  \aap, 465, L35

\bibitem[\protect\citeauthoryear{{Lund}, {Budtz-J{\o}rgensen}, {Westergaard},
  {Brandt}, {Rasmussen}, {Hornstrup}, {Oxborrow}, {Chenevez}, {Jensen},
  {Laursen}, {Andersen} \& {Mogensen}}{{Lund} et~al.}{2003}]{Lund2003}
{Lund} N. et al.,  2003,   \aap, 411, L231

\bibitem[\protect\citeauthoryear{{MacGregor} \& {Vitello}}{{MacGregor} \&
  {Vitello}}{1982}]{MacGregor1982}
{MacGregor} K.~B.,  {Vitello} P.~A.~J.,  1982, \apj, 259, 267

\bibitem[\protect\citeauthoryear{{Martins}, {Schaerer} \& {Hillier}}{{Martins}
  et~al.}{2005}]{Martins2005}
{Martins} F.,  {Schaerer} D.,    {Hillier} D.~J.,  2005, \aap, 436, 1049

\bibitem[\protect\citeauthoryear{{Parmar}, {White}, {Stella}, {Izzo} \&
  {Ferri}}{{Parmar} et~al.}{1989}]{Parmar1989}
{Parmar} A.~N.,  {White} N.~E.,  {Stella} L.,  {Izzo} C.,    {Ferri} P.,  1989,
  \apj, 338, 359

\bibitem[\protect\citeauthoryear{{Perna}, {Bozzo} \& {Stella}}{{Perna}
  et~al.}{2006}]{Perna2006}
{Perna} R.,  {Bozzo} E.,    {Stella} L.,  2006, \apj, 639, 363

\bibitem[\protect\citeauthoryear{{Rahoui}, {Chaty}, {Lagage} \&
  {Pantin}}{{Rahoui} et~al.}{2008}]{Rahoui2008}
{Rahoui} F.,  {Chaty} S.,  {Lagage} P.,    {Pantin} E.,  2008, \aap, 484, 801

\bibitem[\protect\citeauthoryear{{Romano}, {Sidoli}, {Cusumano}, {La Parola},
  {Vercellone}, {Pagani}, {Ducci}, {Mangano}, {Cummings}, {Krimm}, {Guidorzi},
  {Kennea}, {Hoversten}, {Burrows} \& {Gehrels}}{{Romano}
  et~al.}{2009}]{Romano2009}
{Romano} P. et al.,  2009, \mnras, 399, 2021

\bibitem[\protect\citeauthoryear{{Romano}, {Sidoli}, {Ducci}, {Cusumano}, {La
  Parola}, {Pagani}, {Page}, {Kennea}, {Burrows}, {Gehrels}, {Sguera} \&
  {Bazzano}}{{Romano} et~al.}{2010}]{Romano2010}
{Romano} P. et al.,   2010, \mnras, 401, 1564

\bibitem[\protect\citeauthoryear{{Sguera}, {Barlow}, {Bird}, {Clark}, {Dean},
  {Hill}, {Moran}, {Shaw}, {Willis}, {Bazzano}, {Ubertini} \&
  {Malizia}}{{Sguera} et~al.}{2005}]{Sguera2005}
{Sguera} V. et al.,   2005, \aap, 444, 221

\bibitem[\protect\citeauthoryear{{Sguera}, {Bassani}, {Landi}, {Bazzano},
  {Bird}, {Dean}, {Malizia}, {Masetti} \& {Ubertini}}{{Sguera}
  et~al.}{2008}]{Sguera2008}
{Sguera} V. et al.,   2008, \aap, 487, 619

\bibitem[\protect\citeauthoryear{{Shapiro} \& {Lightman}}{{Shapiro} \&
  {Lightman}}{1976}]{Shapiro1976}
{Shapiro} S.~L.,  {Lightman} A.~P.,  1976, \apj, 204, 555

\bibitem[\protect\citeauthoryear{{Shimada}, {Ito}, {Hirata} \&
  {Horaguchi}}{{Shimada} et~al.}{1994}]{Shimada1994}
{Shimada} M.~R.,  {Ito} M.,  {Hirata} B.,    {Horaguchi} T.,  1994, in {Balona}
  L.~A.,  {Henrichs} H.~F.,   {Le Contel} J.~M.,  eds, Proc. IAU Symp. 162, Pulsation; Rotation; and
  Mass Loss in Early-Type Stars. Kluwer, Durdecht, p.487 


\bibitem[\protect\citeauthoryear{{Sidoli}, {Romano}, {Mereghetti}, {Paizis},
  {Vercellone}, {Mangano} \& {G{\"o}tz}}{{Sidoli} et~al.}{2007}]{Sidoli2007}
{Sidoli} L.,  {Romano} P.,  {Mereghetti} S.,  {Paizis} A.,  {Vercellone} S.,
  {Mangano} V.,    {G{\"o}tz} D.,  2007, \aap, 476, 1307


\bibitem[\protect\citeauthoryear{{Sidoli}, {Romano}, {Mangano}, {Pellizzoni},
  {Kennea}, {Cusumano}, {Vercellone}, {Paizis}, {Burrows} \&
  {Gehrels}}{{Sidoli} et~al.}{2008}]{Sidoli2008}
{Sidoli} L. et al.,  2008, \apj, 687, 1230


\bibitem[\protect\citeauthoryear{{Smith}, {Heindl}, {Markwardt}, {Swank},
  {Negueruela}, {Harrison} \& {Huss}}{{Smith} et~al.}{2006}]{Smith2006}
{Smith} D.~M.,  {Heindl} W.~A.,  {Markwardt} C.~B.,  {Swank} J.~H.,
  {Negueruela} I.,  {Harrison} T.~E.,    {Huss} L.,  2006, \apj, 638, 974

\bibitem[\protect\citeauthoryear{{Stevens}}{{Stevens}}{1991}]{Stevens1991}
{Stevens} I.~R.,  1991, \apj, 379, 310

\bibitem[\protect\citeauthoryear{{Stevens} \& {Kallman}}{{Stevens} \&
  {Kallman}}{1990}]{Stevens1990}
{Stevens} I.~R.,  {Kallman} T.~R.,  1990, \apj, 365, 321

\bibitem[\protect\citeauthoryear{{Taam}, {Brown} \& {Fryxell}}{{Taam}
  et~al.}{1988}]{Taam1988}
{Taam} R.~E.,  {Brown} D.~A.,    {Fryxell} B.~A.,  1988, \apjl, 331, L117

\bibitem[\protect\citeauthoryear{{Tarter}, {Tucker} \& {Salpeter}}{{Tarter}
  et~al.}{1969}]{Tarter1969}
{Tarter} C.~B.,  {Tucker} W.~H.,    {Salpeter} E.~E.,  1969, \apj, 156, 943

\bibitem[\protect\citeauthoryear{{Ubertini}, {Lebrun}, {Di Cocco}, {Bazzano},
  {Bird}, {Broenstad}, {Goldwurm}, {La Rosa}, {Labanti}, {Laurent}, {Mirabel}
  \& {Quadrini}}{{Ubertini} et~al.}{2003}]{Ubertini2003}
{Ubertini} P. et al.,   2003, \aap, 411, L131

\bibitem[\protect\citeauthoryear{{Vacca}, {Garmany} \& {Shull}}{{Vacca}
  et~al.}{1996}]{Vacca1996}
{Vacca} W.~D.,  {Garmany} C.~D.,    {Shull} J.~M.,  1996, \apj, 460, 914

\bibitem[\protect\citeauthoryear{{Vedrenne}, {Roques}, {Sch{\"o}nfelder},
  {Mandrou}, {Lichti}, {von Kienlin}, {Cordier}, {Schanne}, {Kn{\"o}dlseder},
  {Skinner}, {Jean}, {Sanchez} \& {Caraveo}}{{Vedrenne}
  et~al.}{2003}]{Vedrenne2003}
{Vedrenne} G. et al.,   2003, \aap, 411,  L63

\bibitem[\protect\citeauthoryear{{Vink}, {de Koter} \& {Lamers}}{{Vink}
  et~al.}{2000}]{Vink2000}
{Vink} J.~S.,  {de Koter} A.,    {Lamers} H.~J.~G.~L.~M.,  2000, \aap, 362, 295

\bibitem[\protect\citeauthoryear{{Wang}}{{Wang}}{1981}]{Wang1981}
{Wang} Y.,  1981, \aap, 102, 36

\end{thebibliography}

\bsp

\label{lastpage}

\end{document}